\documentclass[fleqn,usenatbib]{mnras}

\usepackage{newtxtext,newtxmath}
\usepackage{amsmath}
\usepackage[T1]{fontenc}

\DeclareRobustCommand{\VAN}[3]{#2}
\let\VANthebibliography\thebibliography
\def\thebibliography{\DeclareRobustCommand{\VAN}[3]{##3}\VANthebibliography}

\usepackage{graphicx}	% Including figure files
\usepackage{amsmath}	% Advanced maths commands
\def\2mass0916{2MASS~J0916$-$4215}
\title[A cool, magnetic white dwarf]{A cool, magnetic white dwarf accreting planetary debris}

\author[Vennes et al.]{
St\'ephane Vennes, $^{1}$\thanks{E-mail: svennes@iinet.net.au}
Adela Kawka, $^{2}$
Beth L. Klein, $^{3}$
B. Zuckerman, $^{3}$
Alycia J. Weinberger, $^{4}$
\newauthor
and Carl Melis $^{5}$\\
\\
$^{1}$Mathematical Sciences Institute, The Australian National University, ACT 0200, Australia\\
$^{2}$ International Centre for Radio Astronomy Research - Curtin University, GPO Box U1987, Perth, WA 6845, Australia\\
$^{3}$Department of Physics and Astronomy, University of California, Los Angeles, CA 90095-1562, USA \\
$^{4}$ Earth and Planets Laboratory, Carnegie Institution for Science, 5241 Broad Branch Rd NW, Washington, DC 20015, USA\\
$^{5}$ Center for Astrophysics and Space Sciences, University of California, San Diego, CA 92093-0424, USA\\
}

\date{Accepted XXX. Received YYY; in original form ZZZ}

\pubyear{2023}

\begin{document}
\label{firstpage}
\pagerange{\pageref{firstpage}--\pageref{lastpage}}
\maketitle

% Abstract of the paper2mass0916
\begin{abstract}
We present an analysis of spectroscopic data of the cool, highly magnetic and polluted white dwarf \2mass0916. The atmosphere of the white dwarf is dominated by hydrogen, but numerous spectral lines of magnesium, calcium, titanium, chromium, iron, strontium, along with \ion{Li}{i}, \ion{Na}{i}, \ion{Al}{i}, and \ion{K}{i} lines, are found in the incomplete Paschen-Back regime, most visibly, in the case of \ion{Ca}{ii} lines. Extensive new calculations of the Paschen-Back effect in several spectral lines are presented and results of the calculations are tabulated for the \ion{Ca}{ii} H\&K doublet. The abundance pattern shows a large lithium and strontium excess, which may be viewed as a signature of planetary debris akin to Earth's continental crust accreted onto the star, although the scarcity of silicon indicates possible dilution in bulk Earth material. Accurate abundance measurements proved sensitive to the value of the broadening parameter due to collisions with neutral hydrogen ($\Gamma_{\rm \ion{H}{i}}$), particularly in saturated lines such as the resonance lines of \ion{Ca}{i} and \ion{Ca}{ii}. We found that $\Gamma_{\rm \ion{H}{i}}$ if formulated with values from the literature could be overestimated by a factor of 10 in most resonance lines.
\end{abstract}

%%%%%%%%%%%%%%%%%%%%%%%%%%%%%%%%%%%%%%%%%%%%%%%%%%

%%%%%%%%%%%%%%%%% BODY OF PAPER %%%%%%%%%%%%%%%%%%

\section{Introduction}
The spectroscopic analysis of magnetic white dwarfs covers a wide range of field strengths from $10^{-3}$ to $10^3$ MG. The hydrogen and helium line spectra have been extensively modelled over the whole range of field strength \citep{kem1974,sch2014} but difficulties remain in modelling the field geometry. With few exceptions, magnetic white dwarf are assumed to harbor offset dipole fields. The study of trace elements, from lithium to iron and beyond, in magnetic white dwarf spectra is in its infancy. 

Recent studies have tackled the problem of
line formation in the full Paschen-Back regime in high magnetic fields \citep{zha2018,hol2023}, while \citet{kem1975} initiated the first study of the \ion{Ca}{ii} H\&K doublet in the incomplete Paschen-Back regime \citep[see also][]{har2017}, i.e., between the anomalous regime and the full Paschen-Back regime where the lines assume the shape of a simple polarization triplet ($\pi$ and $\sigma_\pm$) with energy separations simply proportional to the magnetic field strength. 

Meaningful results with abundances of various elements from sodium to iron were obtained at relatively low field in the anomalous Zeeman regime \citep{kaw2011,kaw2014,kaw2019}.
In their study of low-field DZ white dwarfs \citet{hol2021}
encountered the line spectrum of potassium in the incomplete Paschen-Back regime and the lithium spectrum in the full regime. The behavior of spectral lines in magnetic white dwarfs is not yet fully understood and new calculations in fields ranging from 1 to 100 MG are clearly warranted.

Heavy element pollution in white dwarf atmospheres is well documented. Restricting the discussion to cool white dwarfs, accretion of external material, irrespective of the source, is the essential mechanism \citep{dup1993}. In hot white dwarfs, where selective radiation pressure plays an important role \citep{cha1995}, the composition of the atmosphere is dictated by local physical conditions. However, in cool convective white dwarfs the composition of the atmosphere offers clues to the nature and composition of the source itself. Debris material from disrupted planetary bodies is the most likely source \citep{zuc2007,jur2014,zuc2018,ver2021} but variations over this theme rely heavily on the accuracy of abundance measurements. 

Based on a large sample of cool, polluted white dwarfs \cite{hol2018} observed a range of  likely sources, from
(Earth's) crust-like to core-like based on key abundance proportions: calcium, iron or magnesium divided by a weighted sum of the three, which typify, respectively, the Earth's crust, core, and mantle.  Interestingly, most objects point towards a mixture of sources which is labelled bulk Earth. Further, studies of oxygen-polluted white dwarfs have demonstrated evidence of the stoichiometric balance of elements in rock-forming oxides in the accreted material, bearing a close resemblance to rocky bodies in the solar system \citep{kle2010,xus2014,doy2023}.
The detection of the light elements, beryllium and lithium, has raised the possibility that such polluted white dwarfs might be used as tracers of spallation environments \citep{kle2021,doy2021}, differentiation into planetary-crust material \citep{hol2021}, and/or windows into the early universe and Big Bang Nucleosynthesis \citep{kai2021}.
 Overall, the study of abundance patterns in white dwarf stars reveals a complicated history of past and present interaction with their circumstellar environment.

In this context, we present new echelle spectra of the magnetic white dwarf \2mass0916 (Section~\ref{observations}) revealing several spectral lines formed in the incomplete Paschen-Back regime in a 11-12~MG magnetic field (Section~\ref{analysis}). \2mass0916 is a new cool, magnetic white dwarf with trace elements in a hydrogen-rich atmosphere. The abundance pattern shows a two orders of magnitude excess in lithium and an overall distribution, with the notable exception of silicon, pointing out to a primordial origin for the material analogous to Earth's continental crust. A discussion and summary are presented in Section 4. We describe in Appendix~\ref{paschen1} our new calculations of the Paschen-Back effect for all spectral lines found in \2mass0916.

\section{Observations}\label{observations}

The star \2mass0916 appears as a white dwarf candidate in the catalogue of \citet{gen2019}.
We first observed \2mass0916 with the MIKE double echelle spectrograph attached to the Magellan 2 - Clay telescope at 
Las Campanas Observatory on UT 2021 December 18 and 19. We set the slit width to 1 arcsec to provide a resolution $R=\lambda/\Delta\lambda=22\,000$ and 28\,000 on the red and blue sides, respectively \citep{ber2003}. We obtained an additional 
spectrum with the MagE (Magellan echellette) spectrograph \citep{mar2008} attached to the Magellan 1 - Baade telescope on UT 2022 March 23. 
We set the slit width to 0.5 arcsec to provide a resolution $R=8000$. The spectra were corrected for telluric absorption using a template provided by the TAPAS database \citep{ber2014} and using the {\sc telluric} routine within the Image Reduction and Analysis Facility (IRAF). \footnote{IRAF is distributed by the National Optical Astronomy
Observatories, which is operated by the Association of Universities for
Research in Astronomy, Inc. (AURA) under cooperative agreement with the National
Science Foundation.} We employ air wavelengths throughout this work.

We collected optical and infrared photometric measurements from the SkyMapper survey \citep{onk2019}, 
Two Micron All Sky Survey \citep[2MASS:][]{skr2006} and Wide-field Infrared Survey Explorer \citep[WISE:][]{cut2013,mar2021}. 
These are listed in Table~\ref{tbl_astrometry} together with photometric and astrometric data from the Gaia Data 
Release 3 \citep{gaia2022}. \2mass0916 is a nearby white dwarf well within the 40~pc sample \citep{gen2019,obr2023}, also known as WD~J091600.94$-$421520.68 (WD~J0916$-$4215) or Gaia DR3 5427528254746168192, and was tentatively classified as a magnetic white dwarf \citep{obr2023}.

\2mass0916 was observed with the Transiting Exoplanet Survey Satellite \citep[TESS:][]{ric2015} in sectors 35, 36 and 62.
The TESS bandpass has a spectral range from 6000 to 10\,000\AA. Sector 35 was observed from 2021 February 9 to March 6, 
sector 36 was observed from 2021 March 7 to April 1 and sector 62 was observed from 2023 February 12 to March 10.

\begin{table}
 \caption{Astrometry and photometry of \2mass0916.}
 \label{tbl_astrometry}
 \begin{tabular}{lcclcc}
  \hline
  \multicolumn{2}{l}{Parameter} & \multicolumn{3}{c}{Measurement} & Ref. \\
  \hline
  \multicolumn{2}{l}{RA (J2000)} & \multicolumn{3}{c}{09$^{\rm h}$16$^{\rm m}$00\fs94} & 1 \\
  \multicolumn{2}{l}{Dec (J2000)} & \multicolumn{3}{c}{-42\degr 15\arcmin 20\farcs68} & 1 \\
  \multicolumn{2}{l}{$\mu_\alpha \cos{\delta}$ (mas yr$^{-1}$)}& \multicolumn{3}{c}{$-24.67\pm0.04$} & 1 \\
  \multicolumn{2}{l}{$\mu_\delta$ (mas yr$^{-1}$)} & \multicolumn{3}{c}{$-203.56\pm0.05$} & 1 \\
  \multicolumn{2}{l}{$\pi$ (mas)} & \multicolumn{3}{c}{$44.35\pm0.04$} & 1 \\
\hline
\multicolumn{6}{c}{Photometry} \\
\hline
Band & Measurement & Ref. & Band & Measurement & Ref. \\
\hline
  $G$   & $16.566\pm0.003$ & 1 & $J$ & $15.307\pm0.064$ & 3 \\
  $G_{BP}$ & $17.003\pm0.010$ & 1 & $H$ & $15.260\pm0.101$ & 3 \\
  $G_{RP}$ & $15.961\pm0.025$ & 1 & $K$ & $14.890\pm0.124$ & 3 \\
  $g$ & $17.039\pm0.050$ & 2 & $W1$ & $14.770\pm0.015$ & 4 \\
  $r$ & $16.666\pm0.036$ & 2 & $W2$ & $14.725\pm0.020$ & 4 \\
  $i$ & $16.311\pm0.017$ & 2 & $W3$ & $<12.363$ & 5 \\
  $z$ & $16.283\pm0.020$ & 2 &      &                  & \\
  \hline
 \end{tabular}\\
References: (1) \citet{gaia2022}; (2) \citet{onk2019}; (3) \citet{skr2006}; (4) \citet{mar2021}; (5) \citet{cut2013}. 
\end{table}

\section{Analysis}\label{analysis}

The high-dispersion spectra show heavy element lines in a cool hydrogen-rich atmosphere. Nominally, the spectrum is classified as DZAH owing to the presence of strong, magnetic-displaced magnesium, calcium and iron lines and a much weaker hydrogen line.
The presence of a weak, Zeeman-split H$\alpha$ line could imply a temperature slightly above 5\,000~K in a hydrogen-dominated atmosphere but it could also imply a higher temperature, e.g., $\approx 8\,000$~K, in a helium-dominated atmosphere. The temperature and gravity may be constrained further with an analysis of the spectral energy distribution (Section~\ref{sec_model}). All spectral lines display a pattern revealing the presence of a surface average magnetic field of approximately 11.3~MG in strength. With other closely related polluted, hydrogen-rich, magnetic white dwarfs \citep{kaw2019} that were classified as DAZH white dwarfs, \2mass0916\ is part of a class of objects accreting material from their circumstellar environment.

Many spectral lines in \2mass0916\ follow an apparent triplet pattern which at this field strength implies the onset of the full Paschen-Back regime. Closer examination reveals a more complex pattern, particularly in the case of the \ion{Ca}{ii} H\&K doublet. The $\pi$ and $\sigma_\pm$ components follow a pattern already described by \citet{kem1975} which transitions between the low-field anomalous Zeeman and the high-field full Paschen-Back triplet pattern with dominant components shifted by $\Delta E=0$ and $\pm \mu_0B$ and vanishing components at $\pm 2\mu_0B$, where $\mu_0=e\hbar/(2m_e c)$ is the Bohr magneton and $B$ is the magnetic field strength (see Appendix~\ref{paschen1}). In addition, each polarization component is itself a doublet. 

\begin{figure}
	\includegraphics[viewport=10 170 580 690, clip, width=\columnwidth]{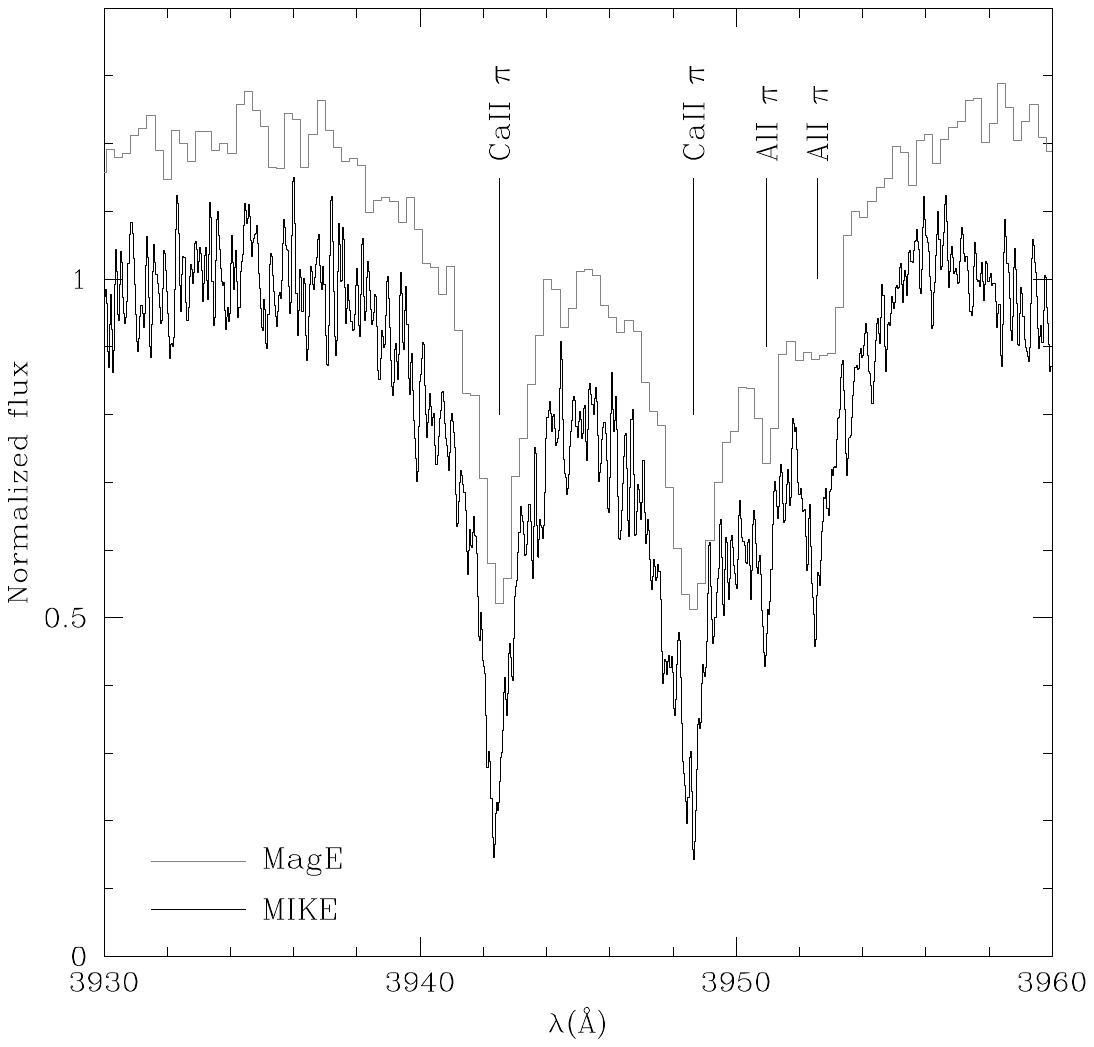}
    \caption{MIKE (black line) and MagE (grey line) echelle spectra showing both \ion{Ca}{ii}~$\lambda$3945 and \ion{Al}{i}~$\lambda3955$ split $\pi$ components in the incomplete Paschen-Back regime.  Wavelengths in all figures are in air.}
    \label{fig_detail_spec}
\end{figure}

This pattern, which can be described as belonging to the incomplete Paschen-Back regime \citep{lan2004}, shows split $\pi$ and $\sigma$ components leaving six strong line components that are characteristic of $^2$S - $^2$P$^{\rm o}$ spectral lines (Fig.~\ref{fig_detail_spec}). For example, sitting close to the $\pi$ components of the \ion{Ca}{ii} H\&K line, split $\pi$ components of the \ion{Al}{i}~$\lambda$3955 line are evident. 
In the Paschen-Back regime the individual polarization components develop a doublet appearance for fine-structure doublets, e.g., \ion{Ca}{ii} H\&K or \ion{Al}{i}~$\lambda$3955, or a triplet appearance for fine-structure triplets, e.g., \ion{Mg}{i}~$\lambda$5178.~\footnote{Throughout the text, to designate spectral lines, we employ the mean multiplet wavelengths obtained from the weighted average (by statistical weight) of the upper and lower energy levels.}
This pattern deviates strongly from the anomalous Zeeman pattern explored by \citet{kaw2019} in the lower field ($\lesssim0.3$~MG) DAZH white dwarf NLTT~7547. The \ion{Ca}{ii} and \ion{Al}{i} lines in a field of $\approx$11.3~MG belong to the incomplete Paschen-Back regime.
 
\subsection{Model atmospheres and stellar parameters}\label{sec_model}

We computed a series of convective model atmospheres containing trace abundance of heavy elements. The electron density is computed self-consistently with the ionization equilibrium of all constituents of the atmosphere including most associated metal monohydrides \citep{tat1984}. 

We note that \citet{bed2017} found that convective energy transfer may be suppressed in a $\approx$10\,000~K hydrogen-rich white dwarf following their spectroscopic analysis of the low-field magnetic white dwarf WD~2105$-$820. Indeed, \citet{tre2015} predicted that even a weak field ($\approx5$~kG) would partially suppress convective energy transfer in the line forming region of an $\approx$10\,000~K hydrogen-rich white dwarf, although vertical mixing may still take place \citep[see also ][]{cun2021}. Similar 3D radiation magnetohydrodynamic calculations applicable to cooler, predominantly neutral atmosphere of cool ($\approx5\,000$~K) DAZ white dwarfs \citep{kaw2019} are not currently available, but setting the plasma-$\beta$ parameter to unity in the line-forming region allows to estimate a critical field of the order of $\approx40$~kG \citep{cun2021}. Although mixing may not be suppressed, the maximum extent of the mixing region under the influence of a strong magnetic field is uncertain but it will be assumed identical to that of a non-magnetic hydrogen envelope (see Section~\ref{sec_diff}). As indicated above, we adopted convective model atmosphere structures.

\begin{figure}
    \centering
    \includegraphics[viewport=10 170 580 690, clip, width=\columnwidth]{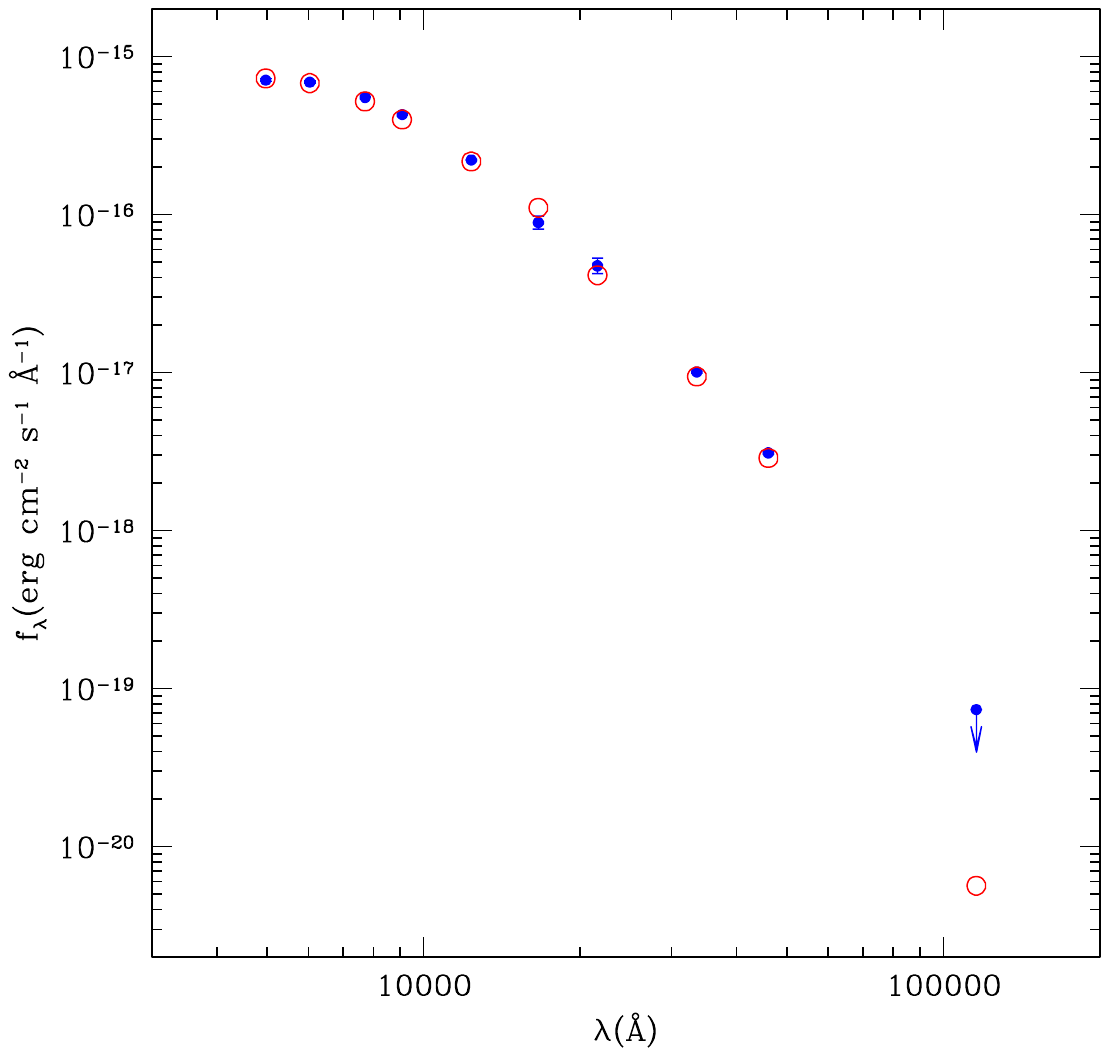}
    \caption{Spectral energy distribution of the cool white dwarf \2mass0916 (blue squares) and best fit synthetic colors (open red circles). The photometric data points are listed in Table~\ref{tbl_astrometry}. We include the Wise W3 band at 12~$\mu$ as an upper limit.}
    \label{fig_sed}
\end{figure}

We first estimated the temperature and surface gravity using all available photometric measurements and the Gaia parallax (Table~\ref{tbl_astrometry}). Figure~\ref{fig_sed} shows a model atmosphere analysis of these measurements: \2mass0916 is a cool white dwarf with
$T_{\rm eff}=5\,250\pm250$~K, and $\log{g} = 8.13\pm0.13$,
where the surface gravity $g$ is expressed in cm\,s$^{-2}$. A weak H$\alpha$ line confirms its low temperature and its hydrogen-rich composition. The strength of the H$\alpha$ $\pi$ component restricts further the range of acceptable parameters with error bars $\sigma(T_{\rm eff})=100$~K and $\sigma(\log{g})=0.06$. We conclude that
\begin{displaymath}
    T_{\rm eff}=5\,250\pm100\,{\rm K},\ {\rm and}\ \log{g} = 8.13\pm0.06
    \end{displaymath}
Using Gaia data, \cite{gen2019} estimated a temperature $T_{\rm eff}=5056$ and 5142~K and a surface gravity of $\log{g}=$7.99 and 8.05, based on pure-helium and pure-hydrogen models, respectively, in agreement with our own measurements. The stellar parameters correspond to a white dwarf mass of $0.66\pm 0.04~M_\odot$ and cooling age of $4.8\pm 0.8$~Gyr. We used the 
evolutionary mass-radius relations of \citet{ben1999}. Note that a slightly longer cooling age of 6~Gyr is obtained with calculations generated by the Montr\'eal White Dwarf Database \citep[MWDD,][]{bed2020} \footnote{\tt https://www.montrealwhitedwarfdatabase.org/evolution.html}.

 \2mass0916 belongs to a class of polluted, cool magnetic white dwarfs originally identified in a high proper-motion surveys such as G~77-50 \citep{far2011}, NLTT~10480 \citep{kaw2011}, NLTT~43806 \citep{zuc2011}, and NLTT~7547 \citep{kaw2019}. \2mass0916 extends the distribution towards higher fields which now covers average surface fields from 70~kG to 11.3~MG.
\subsection{Field strength and structure, and the Paschen-Back effect}\label{field}

The wavelength extent of the triplet line pattern in some lines, e.g., \ion{Mg}{i}~$\lambda5178$, is given by
\begin{equation}
\Delta\lambda (\text{\AA})= 4.67\times10^{-7} \lambda^2(\text{\AA}) B({\rm MG})
\end{equation}
and corresponds to an average surface field of $\approx$11.3~MG.
Despite large wavelength shifts most lines appear narrow. The $\pi$ components are relatively stable in the Paschen-Back regime resulting in minimal broadening when integrating over a dipole field distribution. However the appearance of the $\sigma$ components is sensitive to the field distribution and the offset dipole is a common feature in modelling magnetic white dwarf spectra as it tends to homogenize the field and narrow the spectral lines as observed in \2mass0916.
We computed the field geometry following \citet{mar1981}, \citet{mar1984}, and \citet{ach1989}. An approximate relationship between the dipole field strength $B_{\rm d}$ and its surface average $B_{\rm S}$ is given by
\begin{equation}
    B_{\rm S} \approx 0.7(1+a_z)B_{\rm d}
\end{equation}
where $a_z$ is the offset along the $z$ axis expressed as a fraction of the stellar radius. A centered dipole is located at $a_z=0$. Adopting $a_z=-0.3$ would approximately correspond to an offset dipole field strength of 22~MG close to our dipole field model. The line positions and shapes in the MagE and MIKE spectra are matched with models at $B_{\rm d}=24$~MG and $a_z=-0.3$ inclined at $i=70^{\circ}$. Other examples of offset dipole modelling are presented by \citet{ven2018}.

Other than for the \ion{Ca}{ii} H\&K doublet \citep{kem1975} Paschen-Back calculations are not available in the range of magnetic field reached in \2mass0916.
The \citet{lan2004} theory includes only the linear Zeeman effect and generally applies to low-field stars. In higher field objects such as this one, quadratic field effects as discussed by \citet{kem1975} in the case of the \ion{Ca}{ii} H\&K doublet remain to be explored. As noted earlier by \citet{kaw2019}, both sets of calculations, \citet{kem1975} and \citet{lan2004}, are in agreement at lower fields ($<<$10 MG) where quadratic effects are negligible, although \citet{kaw2019} incorrectly stated that \citet{lan2004} included the effect of quadratic terms. For a general application to spectral lines observed in \2mass0916 it was therefore necessary to develop further the theory elaborated by \citet{lan2004} and include the quadratic terms when computing atomic energy levels embedded in an external magnetic field (Appendix~\ref{paschen1}).

\begin{figure}
	\includegraphics[viewport= 10 170 580 690, clip, width=\columnwidth]{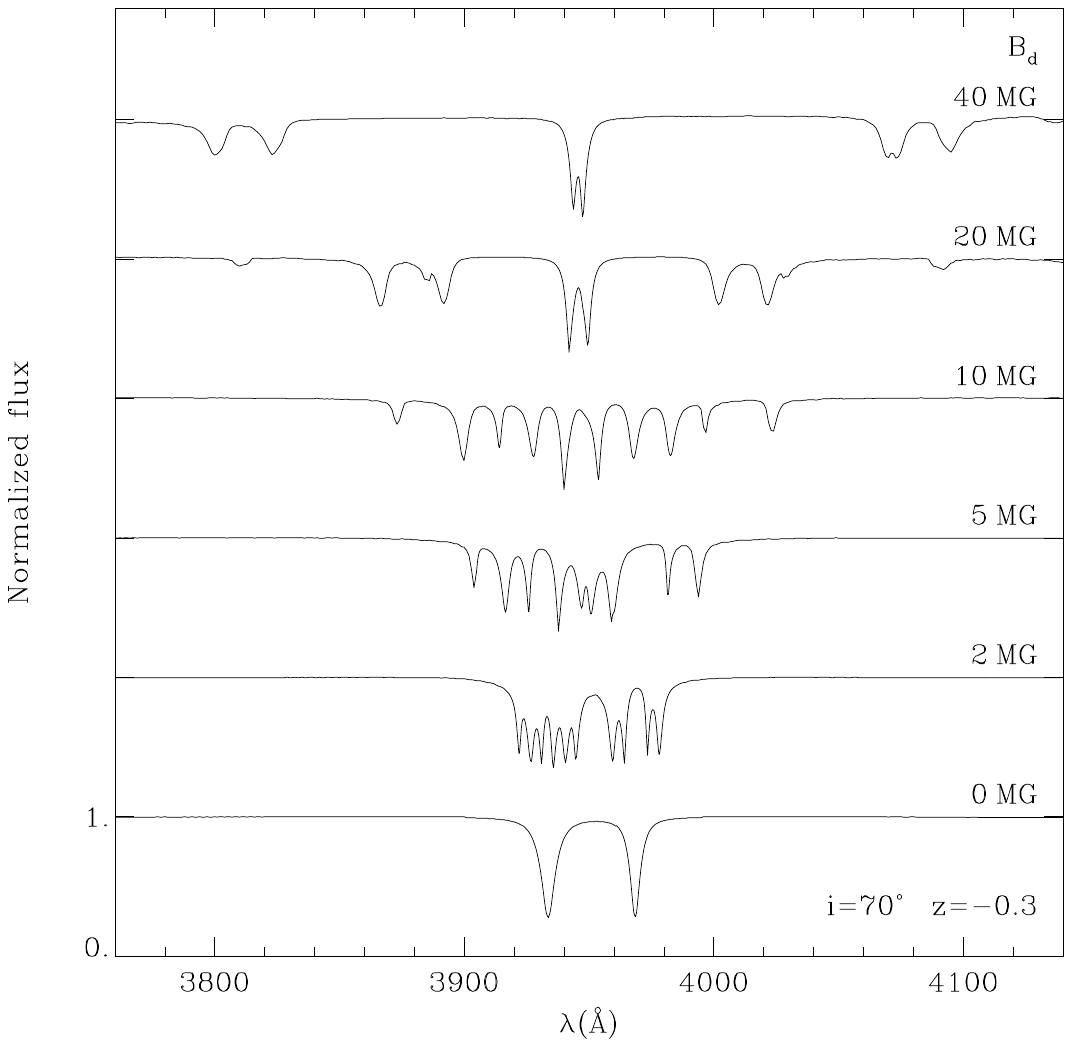}
    \caption{Morphology of the \ion{Ca}{ii} H\&K doublet with increasing dipole field strength ($B_{\rm d}$) from 0 to 40~MG (from bottom to top). The field geometry (inclination and offset) and stellar parameters ($T_{\rm eff}=5\,250$K and $\log{g}=8.13$) of the models are appropriate for the subject of this study, \2mass0916. The transition from the anomalous Zeeman regime (depicted at 2~MG) to the full Paschen-Back regime (depicted at 40~MG) sees the number of visible line components drop from 10 to 6, with the resulting pattern falling in three groups, the central $\pi$ components and the $\sigma_\pm$ components resembling a simple Zeeman triplet. The spectra are shifted vertically by one unit for clarity.}
    \label{fig_models}
    \end{figure}

 Figure~\ref{fig_models} shows a series of \ion{Ca}{ii} H\&K synthetic spectra illustrating the Paschen-Back effect with increasing dipole field strength for a model appropriate for the white dwarf \2mass0916 (see Section~\ref{line_analysis}). The pattern greatly varies with field strength and a model with a dipole field of 20~MG closely resembles the observed calcium spectrum (Fig.~\ref{fig_detail_spec}).
 
\begin{figure}
	\includegraphics[viewport= 10 170 580 690, clip, width=\columnwidth]{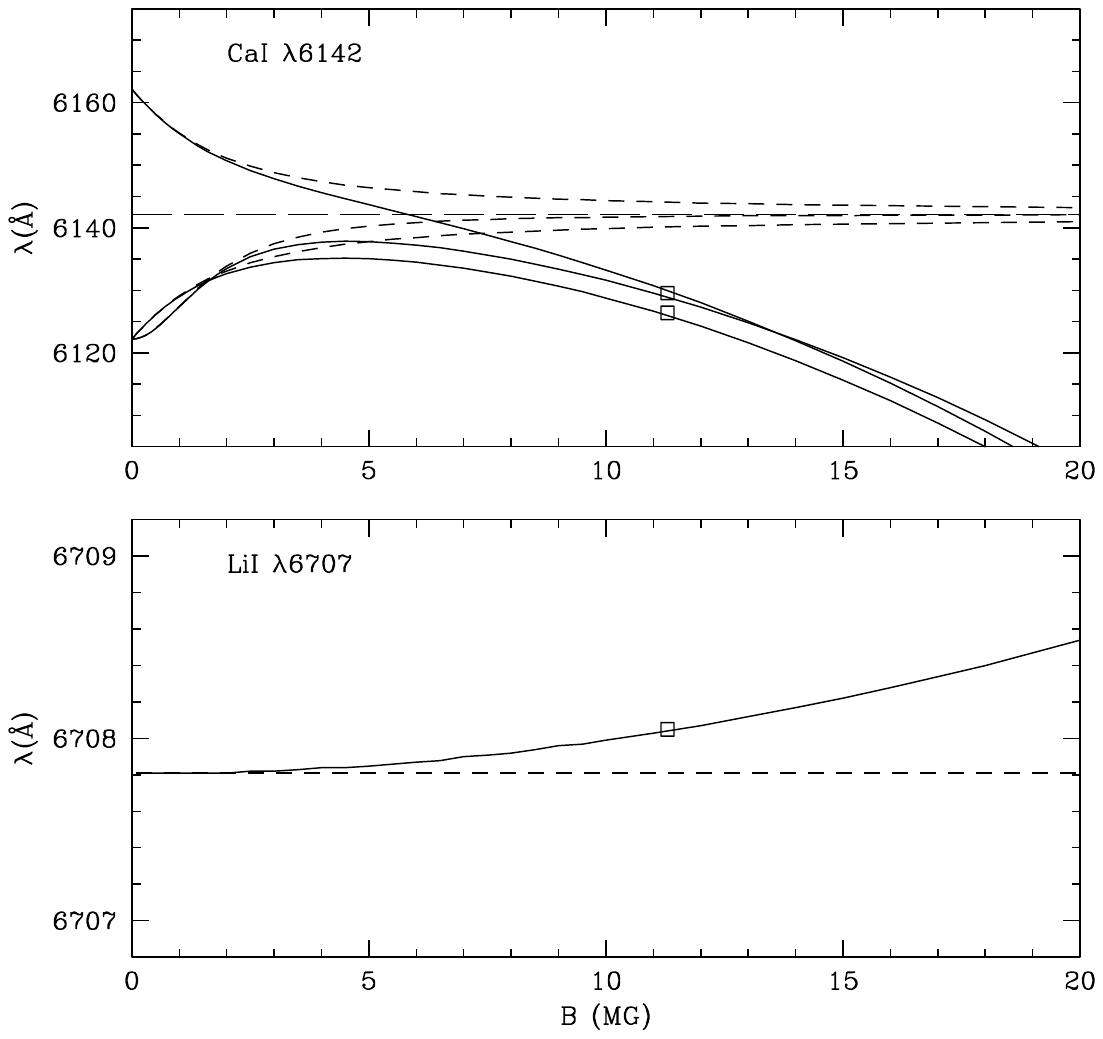}
    \caption{Linear (short dashed lines) and the sum of linear and quadratic (full lines) Paschen-Back calculations compared to observed line positions in the stellar rest frame (open squares, see text). Assuming an average surface field of 11.3~MG, the \ion{Ca}{i}$~\lambda$6142 line is shifted by $\approx-$14\AA\ by the quadratic effect in agreement with the observed line positions, while the \ion{Li}{i}~$\lambda$6707 is shifted by only +0.2\AA. The mean zero-field wavelength of the \ion{Ca}{i} multiplet is shown with a long dashed line.}
    \label{fig_shift}
    \end{figure}

Figure~\ref{fig_shift} shows predicted line positions as a function of magnetic field strength for the singlet \ion{Li}{i}~$\lambda$6707 and the triplet \ion{Ca}{i}~$\lambda$6142. Adopting a field of 11.3~MG corresponding to the average surface field in \2mass0916, we find a quadratic shift of $\Delta \varv=-672.4$~km\,s$^{-1}$ in the $\pi$ component of $\lambda$6142 while it retains a triplet structure, but this shift is only $\Delta \varv=10.4$~km\,s$^{-1}$ in the $\pi$ component of $\lambda$6707. 
Our stellar radial velocity measurement for \2mass0916 is based on a set of these narrow $\pi$ line components. All lines are affected with varying degrees by linear and quadratic shifts. We initially selected \ion{Li}{i}~$\lambda$6707 as the primary velocity indicator because of its small predicted velocity shift. This single line sets the observed heliocentric radial velocity at $\varv_r=21.3$~km\,s$^{-1}$. The gravitational redshift correction will be considered in Section~\ref{kine}. Next, we selected 8 additional lines with increasing Paschen-Back shifts ($\Delta \varv$ in km\,s$^{-1}$) in order to evaluate the reliability of our Paschen-Back models in predicting line shifts: \ion{Ca}{ii}$\lambda$3945 ($\Delta \varv=5.8$), \ion{K}{i}~$\lambda$7676 ($\Delta \varv=-33.1$), \ion{Na}{i}~$\lambda$5891 ($\Delta \varv=-33.7$), \ion{Ca}{i}~$\lambda4226$ ($\Delta \varv=-35.3$), \ion{Sr}{i}~$\lambda$4607 ($\Delta \varv=-44.0$), \ion{Al}{i}~$\lambda$3955 ($\Delta \varv=-281.5$), \ion{Mg}{i}~$\lambda$5178 ($\Delta \varv=-425.0$), \ion{Ca}{i}~$\lambda$6142 ($\Delta \varv=-672.4$). 
The average and dispersion of the radial velocity measurements are $\varv=20.2\pm8$ and excluding two outliers ($\lambda$ 6142 and $\lambda$5178) we find $\Delta \varv=20.6\pm3.7$~km\,s$^{-1}$. We conclude, based on 7 good lines, that  $\varv_r=21\pm4$~km\,s$^{-1}$. Figure~\ref{fig_shift} compares the Paschen-Back calculations to the observed line positions. The predicted $\lambda$6142 line position is in error by $-13$~km\,s$^{-1}$, or $\approx$2\% of the total shift relative to the observed line position (assuming $\varv_r=21$~km\,s$^{-1}$), while the predicted position of the $\lambda$6707 line is in error by $\approx$0.7~km\,s$^{-1}$, or $\approx$7\%. Inclusion of the quadratic shift is not only essential for radial velocity measurements but also in securing correct line identifications. 

\subsection{Line profile and spectral analysis}\label{line_analysis}

The opacity of individual line components is calculated using normalized Lorentzian profiles,
\begin{equation}
    \Phi_\nu = \frac{1}{\pi}\frac{\Gamma/4\pi}{(\nu-\nu_c)^2+(\Gamma/4\pi)^2}
\end{equation}
convoluted with Doppler profiles. Here, $\nu_c$ is the frequency of the shifted line centre, and $\Gamma$ is the sum of the natural width and the broadening parameter, i.e, the full-width at half-maximum expressed in rad\,s$^{-1}$. In cool hydrogen-rich atmosphere, line broadening is dominated by collisions with neutral hydrogen atoms \citep{bar2000b}. 

The hydrogen lines were modelled following the quadratic Zeeman calculations of \citet{sch2014} and using self-broadening parameters of \citet{bar2000a}. We included the linear and quadratic energy corrections in the calculation of the Boltzmann factors for all energy levels. 

The spectra show numerous \ion{Ca}{i} excited lines along with the resonance line \ion{Ca}{i}~$\lambda$4226. The excited lines emerge between 1.9 and 2.5~eV above the ground-state and should only be weakly populated at temperatures near 5\,250~K as the Boltzmann factor for the excited states ranges between $3\times10^{-3}$ and $10^{-2}$ relative to the ground-state. 
\begin{table}
\caption{Calcium broadening parameters at full-width half-maximum $\Gamma/n_{\rm \ion{H}{i}}$ (rad\,s$^{-1}$cm$^3$), $T=5\,250$~K, and population departures.}
    \centering
    \label{table_gamma}
    \setlength\tabcolsep{0.075cm}
    \begin{tabular}{ccccccccc}
    \hline
        &  (1) & (2) && (3) & (4) && (5) & (6) \\
    Ion & $\lambda_{ik}$ & $\log{({\Gamma}/{n_{\rm \ion{H}{i}}})}^a$ && $\log{({\Gamma}/{n_{\rm \ion{H}{i}}})}^b$ & $\log{\rm Ca/H}$ && $\Delta \log{\Gamma}$ & $\log{\rm Ca/H}$\\
             &(\AA)  &        &&         &($+$8.0)&&         &($+$6.7)\\
    \cline{1-3} \cline{5-6} \cline{8-9}
\ion{Ca}{i}  & 4226  &$-$7.88 && $-$7.67 & $+$0.5 && $-$1.0  & $+$0.0 \\
             & 4445  &$-$7.53 && $-$7.26 & $+$1.5 && $-$0.0  & $+$0.0 \\
             & 5266  &$-$7.89 && $-$7.62 & $+$2.0 && $-$0.0  & $+$0.3 \\
             & 5592  &$-$7.88 && $-$7.64 & $+$2.0 && $-$0.0  & $+$0.3 \\
             & 6142  &$-$7.66 && $-$7.30 & $+$1.5 && $-$0.0  & $+$0.0 \\
             & 6460  &$-$7.99 && $-$7.68 & $+$2.0 && $-$0.0  & $+$0.0  \\
 \ion{Ca}{ii}& 3945  &$-$8.08 && $-$7.87 & $+$0.0 && $-$1.5  & $+$0.0 \\
             & 8578  &$-$7.95 && $-$7.78 & $+$1.0 && $-$1.3  & $+$0.0 \\
        \hline
    \end{tabular}\\
      $^a$ Using the \citet{cas2005} formula.
     $^b$ From \cite{bar2000b}.
    \label{tbl_depart}
\end{table}
Abundance measurements in saturated lines, e.g., the \ion{Ca}{ii} H\&K doublet in the spectrum of \2mass0916, are very sensitive to broadening parameter values. On the other hand, abundance measurements are relatively insensitive to broadening parameters in the linear part of the curve-of-growth, e.g., the excited line \ion{Ca}{i}~$\lambda$6142 line in \2mass0916. Detailed line shape measurements obtained with echelle spectra also help in constraining broadening parameter values, e.g., in the \ion{Li}{i}~$\lambda$6707 spectral line. For a simple formalism we estimated the broadening parameter for collision with neutral hydrogen following \citet{cas2005}:
\begin{equation}
\Gamma/n_{\rm \ion{H}{i}} = 4.5\times10^{-9} T_4^{0.3} |\langle r^2_i\rangle-\langle r^2_k\rangle|^{0.4}
\end{equation}
where $T_4=T(K)/10^4$, $\langle r^2_i\rangle$ and $\langle r^2_k\rangle$ are the mean square radii (atomic units) for the lower and upper levels of the transition, respectively, and $n_{\rm \ion{H}{i}}$ is the density of neutral hydrogen. We used as a reference the $\Gamma$ parameters for collisions with neutral hydrogen formulated in \citet{bar2000b} and compared them to values obtained with the \citet{cas2005} formula, both at a temperature of 5\,250~K (3rd and 2nd columns in Table~\ref{table_gamma}). We used mean square radii listed in Appendix~\ref{paschen1} (Tables~\ref{atomic_data} and \ref{data_calcium}).
Values tabulated by \citet{bar2000b} are systematically higher than estimated using the \citet{cas2005} formula but only by a factor of $\approx$2. The width of the strong \ion{Ca}{i}~$\lambda$4300 line calculated using the
\citet{cas2005} formula is exceedingly narrow ($\log{\Gamma/n_{\rm \ion{H}{i}}}\approx-9$). It is not included in the \citet{bar2000b} table.

\begin{table*}
\caption{Abundance of elements ($\log{\rm X/H}$ and $\log{\rm X/Mg}$) by number in \2mass0916 and relative to the bulk Earth, CI-chondrites, and the Earth's continental crust, and estimated mass accretion rate for a given element. Individual errors are estimated at $\pm0.3$~dex.}
\setlength\tabcolsep{0.15cm}
    \centering
    \begin{tabular}{cc ccc ccc ccc c}
    \hline
    Element & $\log{\rm X/H}$ & $\log{\rm X/Mg}$ & $\log{\rm X/Mg}_{\earth}\ ^a$ & &$\Delta_{\rm \earth, Mg}\ ^b$  & & $\log{\rm  X/Mg}_{\rm CI}\ ^c$ & $\Delta_{\rm CI,Mg}\ ^d$ & $\log{\rm  X/Mg}_{\rm cc}\ ^e$ & $\Delta_{\rm cc,Mg}\ ^f$ & $\dot{M}_{\rm acc,A}$ (g~s$^{-1}$)\ $^g$\\
    \hline
        Li & $-$9.30   & $-$2.50  & $-$4.60 & & $+$2.10 & & $-$4.26 & $+$1.76 & $-$2.70 & $+$0.20 & $1.8\times10^5$\\
        Na & $-$6.10   & $+$0.70  & $-$1.91 & & $+$2.61 & & $-$1.25 & $+$1.95 & $-$0.07 & $+$0.77 & $1.3\times10^9$\\
        Mg & $-$6.80   & $+$0.00  & $+$0.00 & & $+$0.00 & & $+$0.00 & $+$0.00 & $+$0.00 & $+$0.00 & $2.9\times10^8$\\
        Al & $-$7.10   & $-$0.30  & $-$1.03 & & $+$0.73 & & $-$1.10 & $+$0.80 & $+$0.43 & $-$0.73 & $2.0\times10^8$\\
        Si & $<-$7.1   & $<-$0.3  & $-$0.04 & & $<-$0.3 & & $-$0.01 & $<-$0.3 & $+$0.94 & $<-$1.2 & $<2.2\times10^8$\\
        K  & $-$7.60   & $-$0.80  & $-$3.19 & & $+$2.39 & & $-$2.45 & $+$1.65 & $-$0.48 & $-$0.32 & $1.1\times10^8$\\
        Ca & $-$6.70   & $+$0.10  & $-$1.17 & & $+$1.27 & & $-$1.25 & $+$1.35 & $+$0.00 & $+$0.11 & $8.2\times10^8$\\
        Ti & $-$8.30   & $-$1.50  & $-$2.57 & & $+$1.07 & & $-$2.62 & $+$1.12 & $-$1.11 & $-$0.39 & $2.6\times10^7$\\
        V  & $<-$8.6   & $<-$1.8  & $-$3.49& & $<$\ 1.7 & & $-$3.57 & $<$\ 1.8 & $-$2.63 & $<$\ 0.8 & $<1.4\times10^7$ \\
        Cr & $-$8.50   & $-$1.70  & $-$1.85 & & $+$0.15 & & $-$1.89 & $+$0.19 & $-$2.65 & $+$0.95 &$1.8\times10^7$ \\
        Mn & $<-$9.0   & $<-$2.2  & $-$2.64 & & $<\ $0.4 & & $-$2.06 & $<-$0.1 & $-$1.91 & $<-$0.3 & $<6.6\times10^6$\\
        Fe & $-$7.00   & $-$0.20  & $-$0.04 & & $-$0.16 & & $-$0.07 & $-$0.13 & $-$0.13 & $-$0.08 & $7.1\times10^8$\\
        Ni & $<-$8.3   & $<-$1.5  & $-$1.31 & & $<-$0.2 & & $-$1.32 & $<-$0.2 & $-$3.06&  $<$\ 1.6 &  $<4.0\times10^7$\\
        Sr & $-$9.35   & $-$2.55  & $-$4.63 & & $+$2.08 & & $-$4.64 & $+$2.09 & $-$2.50 & $-$0.05 & $8.3\times10^6$\\
        \hline
    \end{tabular}\\
    $^a$ Bulk Earth \citep{mcd2003}.
    $^b$ Abundance relative to the bulk Earth: $\log{\rm X/Mg}-\log{\rm X/Mg}_{\rm \earth}$.
    $^c$~CI-chondrites \citep{lod2019}.
    $^d$ Abundance relative to CI-chondrites: $\log{\rm X/Mg}-\log{\rm X/Mg}_{\rm CI}$.
    $^e$ Earth's (bulk) continental crust \citep{rud2003}.
    $^f$~Abundance relative to Earth's (bulk) continental crust: $\log{\rm X/Mg}-\log{\rm X/Mg}_{\rm cc}$.
    $^g$ Mass accretion rate onto the star for a given element in steady-state equilibrium (see Section~\ref{sec_diff}).
    \label{tbl_abun}
\end{table*}
Adopting a model atmosphere at $T_{\rm eff}=5\,250$~K and $\log{g}=8.13$, we computed the line profiles in the Paschen-Back regime described in Appendix~\ref{paschen1} using the broadening parameters for collision with neutral hydrogen listed in \citet{bar2000b}, when available, or calculated using the \citet{cas2005} formula, and fitted them to the MagE and MIKE spectra with a varying abundance. Two problems arose: narrow spectral lines such as \ion{Li}{i}~$\lambda$6707 appeared excessively broadened in the models, and the calcium abundance varied by up to a factor of 100 between measurements based on ground-state lines and measurements based on excited lines (column 4 in Table~\ref{tbl_depart}). Reductions in $\Gamma$ values of up to 1.5 orders of magnitude ($-1.5$~dex; column 5) were necessary to reconcile calcium abundance measurements based on individual lines (column 4). As expected, saturated ground-state lines required large reductions in $\Gamma$ values while unsaturated  excited lines required no corrections. The resulting abundance measurements are 1.3~dex higher, i.e., $\log{\rm Ca/H}=-6.7$, than estimated using un-corrected $\Gamma$ values, i.e., $\log{\rm Ca/H}=-8.0$, and are mutually consistent (column 6).

 Similar difficulties arose in two separate measurements of the sodium abundance: adopting the reference parameters $\Gamma$ the abundance based on the strong resonance line \ion{Na}{i}~$\lambda$5891 is one order of magnitude lower than the abundance measured using the weak excited line \ion{Na}{i}~$\lambda$8190. By adjusting the broadening parameter $\Gamma$ for the $\lambda$5891 line by $-1.7$~dex we obtained consistent abundance measurements at $\log{\rm Na/H}=-6.1$.

Finally, to resolve difficulties in matching the synthetic line profiles to the observed ones, literature-based $\Gamma$ values for the narrow ground-state lines \ion{Li}{i}~$\lambda$6707, \ion{Al}{i}~$\lambda$3955, \ion{K}{i}~$\lambda$7676, and \ion{Sr}{i}~$\lambda$5607 were reduced by $1.0$ dex. Similarly a modest correction of $-0.5$~dex was required in the case of \ion{Mg}{i}~$\lambda$5178 to match its narrow width. The resulting abundance measurements based on unsaturated spectral lines were not affected by this procedure. We adopted a correction of $-1.0$~dex for the remaining line broadening parameters.
Our echelle spectra exposed the need for considerable revisions in broadening parameter values. \citet{hol2021} and \citet{elm2022} also concluded that broadening parameters by collisions with neutral atoms required large correction factors. 

\begin{figure}
	\includegraphics[viewport= 10 170 580 690, clip, width=\columnwidth]{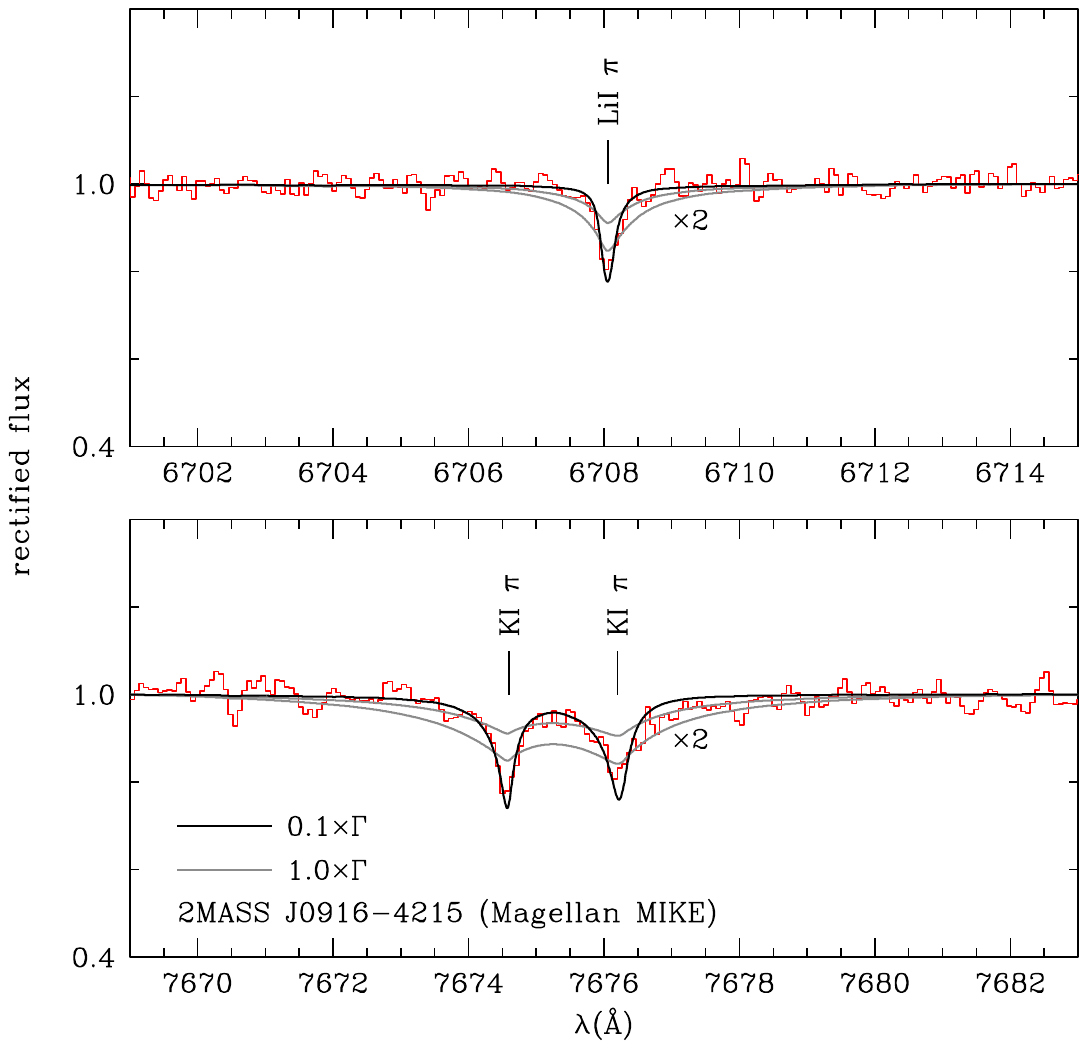}
    \caption{Telluric-corrected MIKE echelle spectrum (red line) for the \ion{Li}{i} line (top panel) and \ion{K}{i} line (bottom panel) compared to model spectra with broadening parameters at full value (grey line) and divided by 10 (black line). We show again the profiles at full width but with an increased abundance ($\times2$). The \ion{K}{i} $\pi$ doublet appearance is a signature of the Paschen-Back effect at a field of 11.3~MG while \ion{Li}{i} $\pi$ achieves a singlet appearance.}
    \label{fig_LiK}
\end{figure}
Interestingly, \citet{kaw2011}, \citet{kaw2014}, and \citet{kaw2019} have already noted that calcium abundance measurements based on \ion{Ca}{i}~$\lambda$4226, \ion{Ca}{ii} H\&K, and the \ion{Ca}{ii}~$\lambda$8578 were often inconsistent. Based on our new results we propose to revise upward the calcium abundance measurement in NLTT~7547 that was presented by \citet{kaw2019} from $\log{\rm Ca/H}=-10.1$ to $-9.3$, i.e, adjusted to the measurement based on the unsaturated \ion{Ca}{ii}~$\lambda$8578 line.

Figure~\ref{fig_LiK} shows our analysis of the \ion{K}{i}~$\lambda$7676 and \ion{Li}{i}~$\lambda$6707 line profiles. The narrow line shapes were matched by synthetic line profiles including, as noted above, reduced broadening parameters ($\log{\Gamma}-1.0$). The line profiles computed at the same abundance but with the original line broadening parameters obtained from \citet{bar2000b} are clearly too broad and shallow with line wings extending far from the line centre. To show this effect more clearly we also show the full profiles at twice the nominal abundance. The line positions are accurately matched with our Paschen-Back models including the quadratic effect. Their singlet (\ion{Li}{i}) and doublet ({\ion{K}{i}}) appearances follow directly from their respective fine-structure energy separation constants $\zeta$ which is much larger in the case of the \ion{K}{i}~$\lambda$7676 upper level ($^2$P$^{\rm o}$) relative to the \ion{Li}{i}~$\lambda$6707 upper level, $\zeta=$38.5 versus 0.2~cm$^{-1}$, or the \ion{Na}{i}~$\lambda$5891 upper level (see Appendix~\ref{paschen2}). 

\begin{figure*}
	\includegraphics[viewport=40 250 590 620, clip, width=17cm]{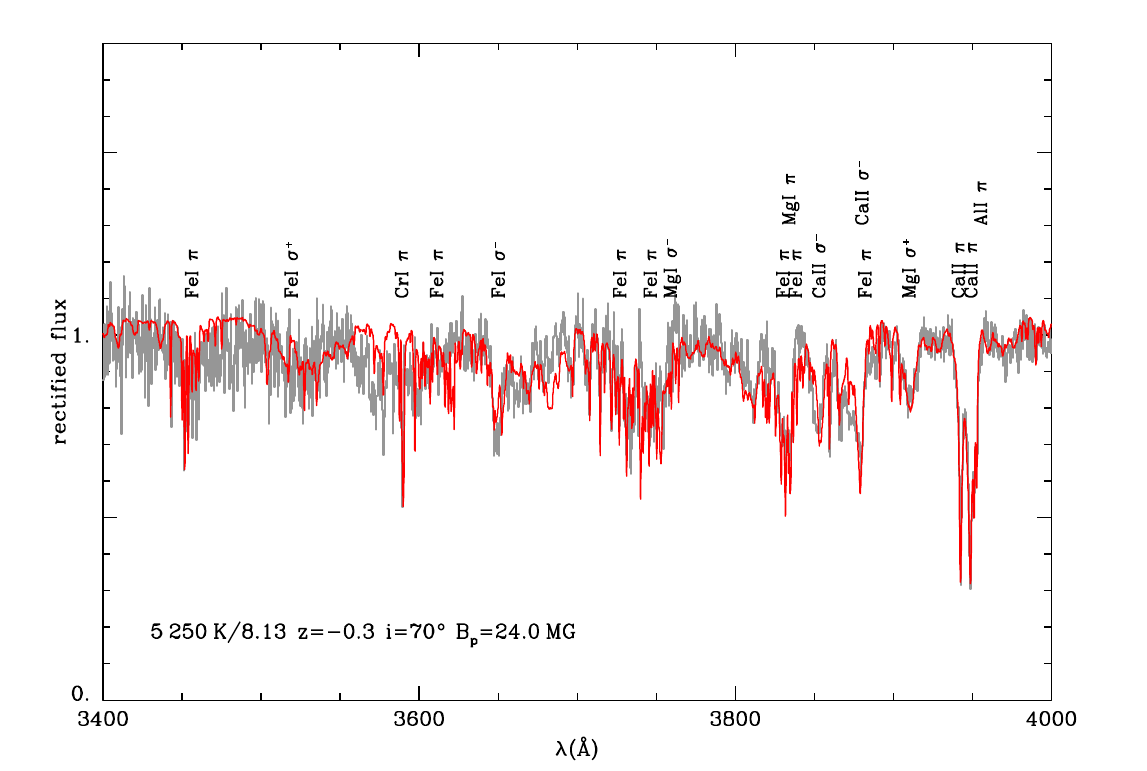}
    \caption{Model spectrum (red line) fitted to the blue side of the MagE echelle spectrum (grey line) using air wavelengths. All spectral lines modelled in this work are marked with the corresponding element and polarization state ($\pi$ or $\sigma_\pm$). The complete spectral coverage is presented in Appendix~\ref{atlas}.} 
    \label{fig_spec}
\end{figure*}

Table~\ref{tbl_abun} lists the resulting abundance measurements by number and Figure~\ref{fig_spec} shows the corresponding spectral synthesis along with the Magellan echellette (MagE) spectrum (see also Appendix~\ref{atlas}). We estimate the individual abundance uncertainties at a factor of 2 ($\pm0.3$~dex). They are dominated by uncertainties in the effective temperature of the star ($\pm100$~K) and to a lesser extent the uncertainties in the broadening parameters, and the line profile fitting. Therefore, because systematic errors dominate the error budget, errors in abundance ratios should be lower than in individual measurements. 

All \ion{Ca}{ii} lines belong to the incomplete Paschen-Back regime. The ultraviolet \ion{Ca}{ii} lines show the close doublet pattern ($\pi$) as well as split $\sigma_\pm$ patterns $\approx \pm85$\AA\ on both sides and, as expected at $\approx11.3$~MG (see Equation (1)), a weak vanishing component at $\Delta \lambda\approx +170$\AA\ ($\lambda \approx4120$\AA). The infrared \ion{Ca}{ii} sextet shows a corresponding number of $\pi$ components (five are clearly visible in the spectrum and a sixth one is merged with another). The $\sigma_\pm$ components of \ion{Mg}{i}~$\lambda$5178 and \ion{Ca}{i}~$\lambda$4445 (a sextet at zero field) show a complex triplet structure well matched by the models. Several narrow lines do not have obvious identifications and are shown with "?" marks. A possible identification of a feature near 4480~\AA\ with the excited \ion{Mg}{ii}~$\lambda$4482 line is problematic because of its high excitation energy and vanishing Boltzmann factor at low temperature.

\begin{figure}
	\includegraphics[viewport=10 170 580 690, clip, width=\columnwidth]{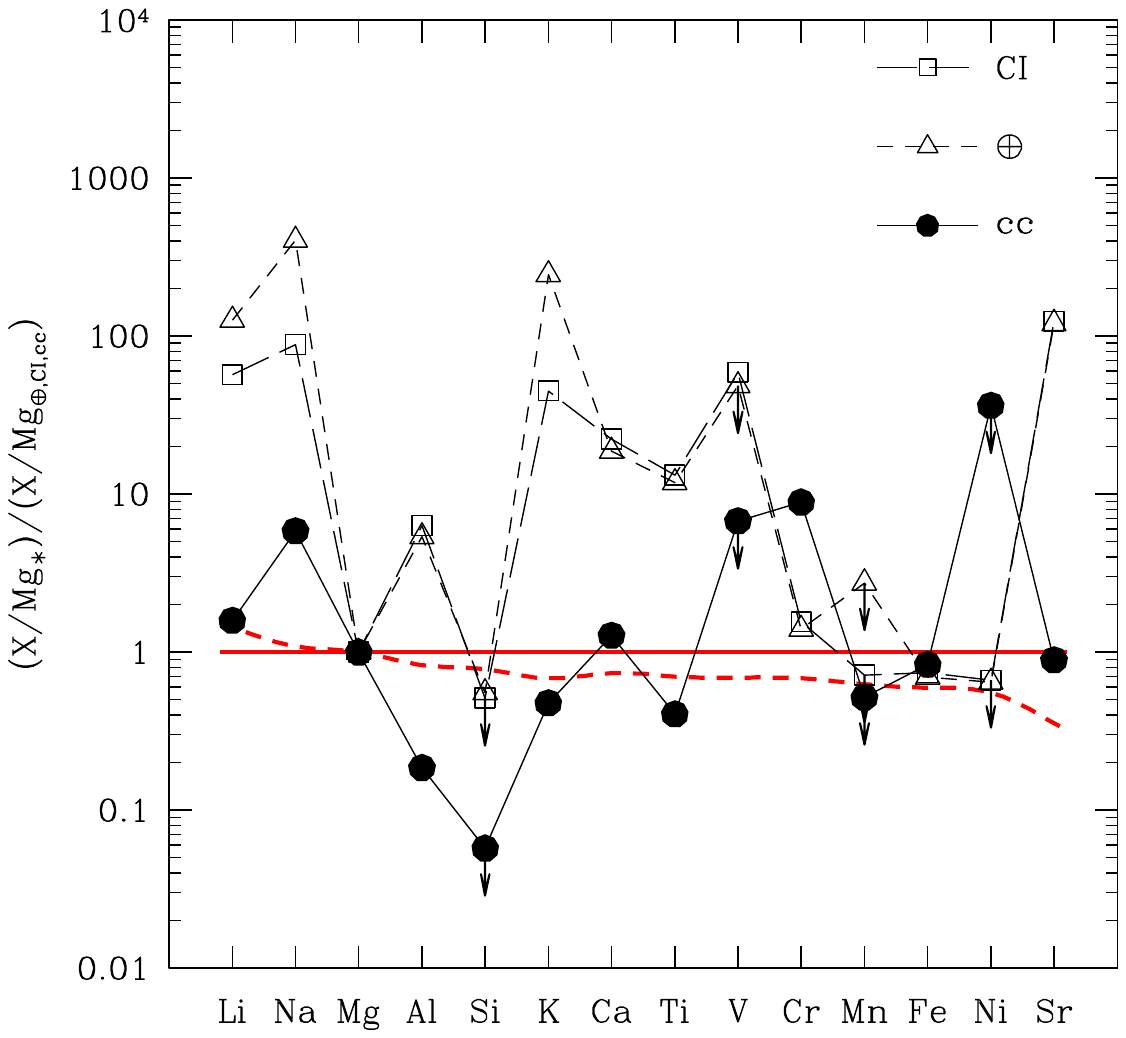}
    \caption{Abundance set in the photosphere of \2mass0916. The abundance measurements (by number) are normalized to magnesium (${\rm X/Mg_\ast}$) and divided by the same measurements in Earth's continental crust (cc), CI-chondrites (CI), and the bulk Earth composition ($\earth$) as listed in Table~\ref{tbl_abun}. Silicon, vanadium, manganese, and nickel abundance upper limits are marked with down arrows. The full red line ($=1$) follows the early build-up assumption, while the dashed red line assumes accretion-diffusion equilibrium (see text).}
    \label{fig_abun}
\end{figure}
Figure~\ref{fig_abun} shows our new abundance measurements listed in Table~\ref{tbl_abun}. The abundance measurements are normalized to the magnesium abundance and divided by the corresponding abundance measurements in various bodies and the resulting ratios are plotted for each element. Excluding upper limits (e.g., silicon), the mean and standard deviation of the ratios relative to Earth's continental crust are $0.05\pm0.51$, while these values increase to  $1.20\pm0.73$ relative to the CI-chondrites and $1.36\pm0.94$ relative to the bulk Earth.

The lithium over-abundance in \2mass0916 approaches $+2$~dex relative to the bulk Earth \citep{mcd2003} or CI-chondrites \citep{lod2019}, but it appears considerably closer to the lithium content of Earth's (bulk) continental  crust \citep{rud2003}. The strontium abundance would also point towards a parent body composed of crust-like material. However, the upper limit on the abundance of silicon would imply a large silicon deficit in the parent body assuming Earth's crust composition. On the other hand, the upper limits on the abundance of vanadium, manganese, and nickel do not help discriminate between the three scenarios.
The standard deviation in $\Delta_{\rm cc,Mg}$ measurements is $\approx0.5$ showing an uncertainty in individual measurements of the order of a factor of 3, or some degrees of variations in the  composition of the actual accreted material relative to the estimated composition of crust-like material \citep{rud2003}.
Oxygen is not detectable in cool white dwarfs such as \2mass0916, so possible oxide-balance of the accreted material cannot be ascertained \citep{kle2010}.

\subsection{Effect of diffusion: build-up and steady-state regimes}\label{sec_diff}

Before any chemical separation takes effect, i.e., at a time following an accretion event much shorter than the diffusion time scale, $t<<\tau$, the abundance pattern (relative to Mg) in the atmosphere replicates the pattern in the accretion flow:
\begin{equation}
    \Big{(}\frac{\rm X}{\rm Mg}\Big{)}_{\ast}=\Big{(}\frac{\rm X}{\rm Mg}\Big{)}_{\rm acc} 
\end{equation}
The comparisons with the full red line depicted in Figure~\ref{fig_abun} follow this assumption.
However, in a steady-state regime at a time $t>>\tau$ when equilibrium is established between diffusion losses at the bottom of the convection zone and the surface resupply, the abundance pattern (relative to Mg) is given by:
\begin{equation}
    \Big{(}\frac{\rm X}{\rm Mg}\Big{)}_{\ast}=\Big{(}\frac{\rm X}{\rm Mg}\Big{)}_{\rm acc} \  \frac{\rm \tau(X)}{\rm \tau(Mg)}
\end{equation}
where $\tau{\rm (X)}$ is the diffusion time scale for a given element X (or specifically Mg) obtained from the MWDD \citep{bed2020}. We note that the mass of the convection zone is calculated assuming that the strong magnetic field has no effect on the depth of the mixed layers in a magnetic white dwarf.
The dashed red line in Figure~\ref{fig_abun} shows mild deviations in the abundance pattern due to diffusion. Assuming either early build-up or steady-state regimes, accretion from a source with crust-like material appears more likely. However, the absence of silicon remains problematic. Moreover, sodium, aluminium and chromium deviate from expected (either in build-up phase or steady-state phase) abundance ratios by more than 0.5~dex. Such large deviations exceed mere statistical errors and may reflect systematic errors in adopted broadening parameter values.

\begin{figure*}
	\includegraphics[viewport= 10 170 580 690, clip, width=0.7\textwidth]{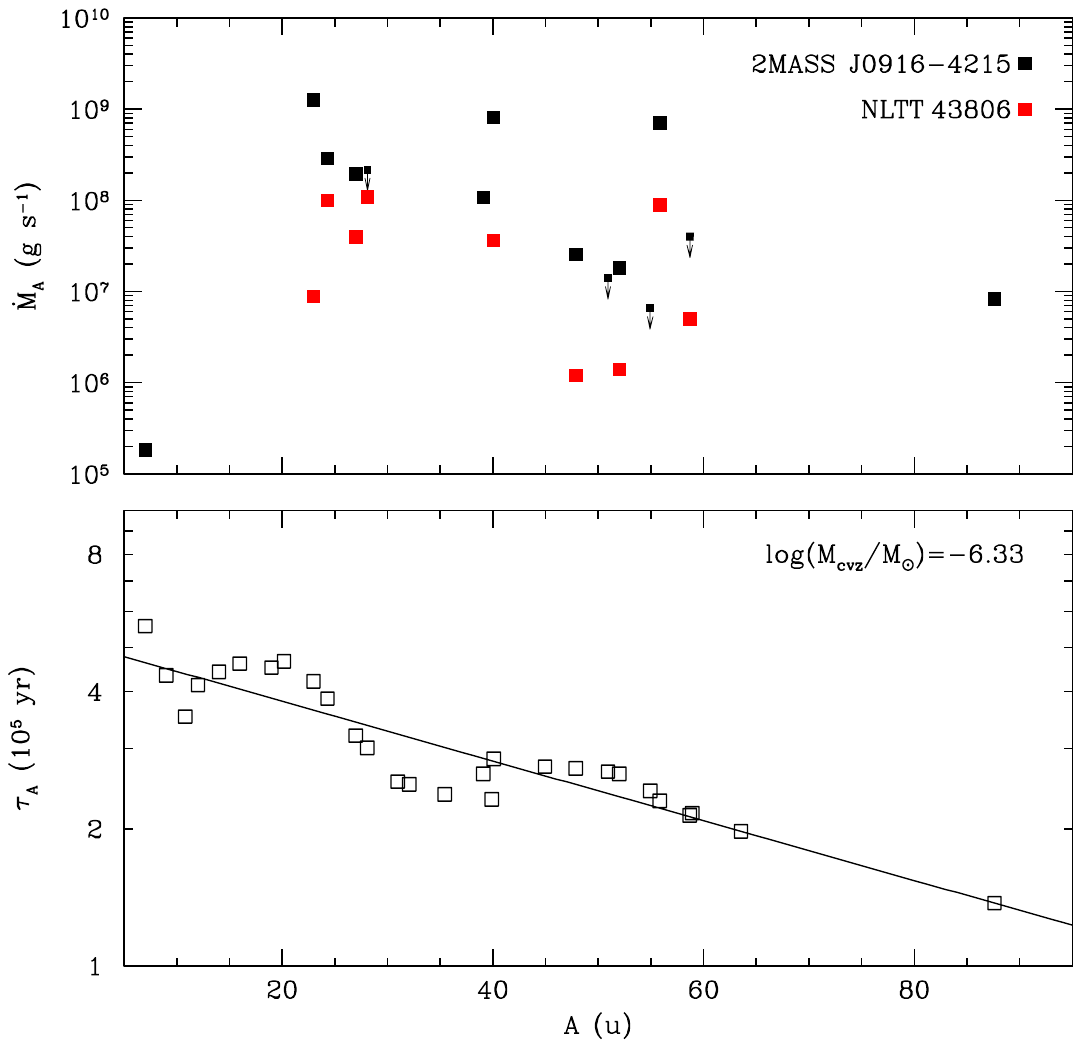}
    \caption{Estimated mass accretion rate (top panel) onto the photosphere of \2mass0916 (black squares) and NLTT~43806 \citep[red squares,][]{zuc2011} assuming steady-state equilibrium. Rate upper limits are marked with down arrows. Diffusion time scales (open squares in bottom panel) for a set of elements with atomic weight from $A=6.9$~u (Li) and $A=63.6$~u (Cu) obtained from the MWDD \citep{bed2020}. $M_{\rm cvz}$ is the adopted mass of the convection zone expressed as a fraction of the solar mass. The diffusion time scale for strontium ($A=87.6$~u) was estimated by extrapolating available data (full line).}
    \label{fig_diff}
\end{figure*}

  The mass of individual elements accreted onto the star per units of time is given as its mass fraction $Y_{\rm acc}$ of the accretion flow and is expressed in $M_\odot\,{\rm yr}^{-1}$ or g~s$^{-1}$:
\begin{equation}
   \dot{M}_{\rm acc,A} = Y_{\rm acc,A}\dot{M}_{\rm acc} = Y_{\rm atm,A}\ \frac{M_{\rm cvz}}{\tau_{\rm A}}
\end{equation}
where $A$ identifies an element by its atomic weight, $\dot{M}_{\rm acc}$ is the total mass accreted by units of time, $Y_{\rm atm, A}$ is the mass fraction of element $A$ in the atmosphere, $Y\approx A\,X$, where $X$ is the abundance by number (${\rm X/H}$ in Table~\ref{tbl_abun}), $M_{\rm cvz}$ is the mass of the convection zone ($\log{\rm M_{cvz}/M_\odot}=-6.33$), and $\tau_{\rm A}$ is the diffusion time-scale of element $A$ at the bottom of the convection zone. Figure~\ref{fig_diff} shows the diffusion time-scales employed in the calculation of individual accretion rates onto \2mass0916. 

The resulting accretion rates for individual elements are compared to similar rates onto the polluted, magnetic white dwarf NLTT~43806 \citep{zuc2011}. The rates found in \2mass0916\ are a factor of 6 larger than in NLTT~43806 but, otherwise, follow a very similar trend. \citet{zuc2011} concluded that the material accreted onto NLTT~43806 belong to lithosphere-like material (crust and upper mantle). Although the overall abundance pattern in \2mass0916\ would suggest a similar conclusion, the scarcity of aluminium and the absence of silicon indicate that the silicon-rich crust material should be diluted with other types of material. We find that, depending on the adopted scenario, i.e., early build-up phase (equation (6)) or late steady-state phase (equation (7)), individual abundance measurements would vary by at most a factor of 3 due to variations in the diffusion time scales, particularly between lighter and heavier elements. Such mild variations are not readily detectable in our measurements. Therefore, we cannot recover the timeline of accretion events onto \2mass0916. 

Complete or partial suppression of convective mixing could noticeably shorten diffusion time-scales in cool white dwarfs. Following an accretion event the fully convective envelope of a 5\,000~K hydrogen-rich white dwarf would retain heavy elements over a time-scale of $\approx10^6$ years, while a radiative photosphere without mixing would retain its composition for merely $\approx10^{-3}$ years. Unless they are in the process of accreting material, the cool, polluted magnetic white dwarfs would be relatively rare compared to their non-magnetic counterparts. However,
an examination of the large sample of polluted, magnetic white dwarfs analyzed by \citet{hol2017} and \citet{hol2018} shows that the abundance patterns of magnetic and non-magnetic white dwarfs do not differ significantly and that magnetic fields are common among polluted white dwarfs which implies that deep mixing remains effective even among magnetic white dwarfs.

\subsection{Kinematics}\label{kine}

Using the distance, proper motion and the stellar radial velocity of $\varv = -15\pm4$~km~s$^{-1}$, obtained by subtracting a gravitational redshift of 36~km~s$^{-1}$ from the observed heliocentric radial velocity $\varv_r = 21$~km~s$^{-1}$ (see Section~\ref{field}), we calculated the Galactic velocity components relative to the local standard of rest following \citet{joh1987}. The velocities were corrected for the Solar motion relative to the local standard of rest using ($U_\odot,V_\odot,W_\odot$) = ($11.10, 12.24,7.25$)~km~s$^{-1}$ \citep{sch2010}. The Galactic velocity components of ($U,V,W$) = ($29.3\pm0.2, 25.2\pm4.0,-7.1\pm0.3$)~km~s$^{-1}$ place \2mass0916 in the thin disc.

\subsection{Photometric variations}

We could not detect any significant photometric variations in \2mass0916 in the TESS data. Several other cool, magnetic white dwarfs show variations of several hundredth of a magnitude, e.g., NLTT~8435 \citep{ven2018}, possibly due to surface field variations or surface elemental abundance variations.  However, no surface element abundance changes other than for hydrogen and helium \citep[see, e.g.,][]{cai2023}, as measured with absorption line equivalent widths, have ever been convincingly demonstrated for any white dwarf \citep[e.g., see Section 3.2 in][]{joh2022}. With its large element abundances and complex but well-modelled line spectrum, \2mass0916 would be the ideal candidate for such a study.

\section{Discussion and Summary}

The intermediate field white dwarf \2mass0916 is among the first of its class to show spectral lines in the incomplete Paschen-Back regime. Previously, this pattern was observed in the \ion{K}{i} spectrum of the lower field DZ white dwarf LHS~2534 \citep{hol2021}. The $\pi$ components of several spectral lines show doublet, e.g., \ion{Al}{i}~$\lambda$3955, \ion{K}{i}~$\lambda$7676, and \ion{Ca}{ii}~$\lambda$3945, or triplet appearances characteristic of the incomplete Paschen-Back regime. The infrared \ion{Ca}{ii}~$\lambda$8578 triplet shows a complex multi-components core structure that falls well into the incomplete Paschen-Back regime. The pattern clearly departs from the anomalous Zeeman effect displayed in several lower-field white dwarfs \citep{kaw2011,kaw2014,kaw2019} but it shows split components and residuals of line components that should vanish entirely in the full Pashen-Back regime. We have presented new calculations of the incomplete Paschen-Back regime and listed sample results for the \ion{Ca}{ii} H\&K doublet. Although we developed a reliable method to adjust broadening parameters in the high density atmosphere of cool white dwarf stars, ab initio calculations of these parameters are needed to help eliminate potential systematic errors in abundance measurements.

To date, lithium has been detected in only seven white dwarfs \citep[][this work]{kai2021,hol2021,elm2022}, including \2mass0916, all of them cooler than $\approx 5000$~K with the exception of \2mass0916. Three of them are magnetic which suggests a high incidence of magnetism in cool, polluted white dwarfs \citep{kaw2014,hol2018,kaw2019}. The lithium to sodium abundance ratio varies from $\log{\rm Li/Na}=-0.1$ to $-3.2$ (in \2mass0916) but with the five other measurements clustering around $-1.8\pm0.3$. The material accreted on \2mass0916 may not be as "differentiated" as the material analyzed by \citet{hol2021}. The low ionization potential of neutral lithium precludes a detection in warmer objects but a search for this element in objects with temperatures up to 6\,000~K should be attempted. 

\2mass0916 joins a class of cool, polluted hydrogen-rich white dwarfs as its highest field member ($B_{\rm S}\approx11.3$~MG, $B_{\rm p}=24$~MG). 
Pending a definitive model atmosphere analysis, SDSS~J1143+6615 \citep{hol2023} could constitute an even higher field member of this class although a hydrogen-rich composition appears unlikely. The spectral energy distribution of \2mass0916 does not show an excess in the infrared, however the WISE upper limit for the 12$\mu$m flux measurement provides for the possibility of an infrared excess in the mid-infrared range. JWST mid-infrared imaging is needed to investigate the presence of a dusty circumstellar environment. The overall abundance pattern indicates a lithium- and strontium-rich source of material similar to the Earth's crust but the scarcity of aluminium and silicon argues against this simple interpretation and qualitatively different sources of material are also required. 

\section*{Acknowledgements}
C.M.\ and B.Z.\ acknowledge support from US National Science Foundation
grants SPG-1826583 and SPG-1826550.
%%%%%%%%%%%%%%%%%%%%%%%%%%%%%%%%%%%%%%%%%%%%%%%%%%
\section*{Data Availability}
The MagE and MIKE echelle spectra may be obtained from the authors upon request. These data sets are not currently available on public archives.


%%%%%%%%%%%%%%%%%%%%%%%%%%%%%%%%%%%%%%%%%%%%%%%%%%

%%%%%%%%%%%%%%%%% APPENDICES %%%%%%%%%%%%%%%%%%%%%

\appendix

\section{Incomplete Paschen-Back regime}\label{paschen1}

We revise and expand upon the original calculations presented by \citet{kem1975} using methods presented by \citet{cow1981}, \citet{gri1995}, and \citet{lan2004}.
In Section~\ref{calcium} we present our calculations of line strengths and positions for the Ca{\small II} H\&K doublet components under the incomplete Paschen-Back regime, and in Section~\ref{paschen2} we present new results for other spectral lines of interest under the same regime.

We tabulate relevant input data for our calculations in Tables~\ref{atomic_data} and ~\ref{data_calcium}, with the assistance of the NIST Atomic Spectra Database \citep{kra2022}. Results of our calculations (line position and strength) for the \ion{Ca}{ii} H\&K doublet are presented in Tables~\ref{tbl_CaHK_S} and ~\ref{tbl_CaHK_W}.

\subsection{The Ca~{\small II} H\&K doublet}\label{calcium}

\citet{kem1975} wrote the Hamiltonian for an atom embedded in an external magnetic field as the sum of the spin-orbit, linear Zeeman and quadratic terms:
\begin{equation}
{\rm H} = {\rm H}_{\rm so}+\beta\,{\rm H}_{\rm B} + \gamma B^2\,{\rm H}_{\rm Q}
\end{equation}
where $\beta=\mu_0B$ and $\mu_0=e\hbar/(2m_e c)$ is the Bohr magneton acting at a magnetic field strength $B$, and $\gamma=e^2/(8m_e c^2)$ is a constant factor applied to the quadratic term ${\rm H_Q}$, where all constants have their usual meaning.

In the matrix below, we show the linear Paschen-Back (H$_{\rm B}$) matrix elements added to the spin-orbit (H$_{\rm so}$) elements for the upper energy levels (4p $^2$P) of the Ca{\small II} H\&K doublet. The matrix elements due to $H_B$ are $g_L M\beta$, where $g_L=g_L(J,L,S)$ is the Lande $g$ value and $M$ the magnetic quantum number, while the matrix elements due to $H_{\rm so}$ describe the level fine-structure. The formulation recovers the anomalous Zeeman effect in the low field limit $\beta << \zeta$, e.g., $B\lesssim 0.5$\,MG for Ca{\small II} H\&K. Here, $\zeta$ is the energy separation constant for a given multiplet, e.g., $\zeta=148.593$\,cm$^{-1}$ for the Ca{\small II} 4p configuration. The resulting tri-diagonal matrix structure, $(J,J)$ on the diagonal and, when allowed, $(J,J-1)$ and $(J-1,J)$ off the diagonal, shows the mixed $J$ levels with a common $M$ number at $M=+1/2$ (lines 3 and 4) and $M=-1/2$ (lines 5 and 6) which alters the level structure at higher field:
\begin{displaymath}
\mathbf{\rm H_{so}+H_B} =
\end{displaymath}
\begin{displaymath}
\left( \begin{array}{cccccc}
\frac{\zeta}{2}+2\beta & 0 & 0 & 0 & 0 &0  \\
   0     & \frac{\zeta}{2}-2\beta & 0 & 0 & 0 & 0 \\
   0     & 0 & \frac{\zeta}{2}+\frac{2\beta}{3} & -\frac{\sqrt{2}}{3}\beta & 0 & 0 \\
  0      & 0 & -\frac{\sqrt{2}}{3}\beta & -\zeta+\frac{\beta}{3} & 0 & 0 \\
     0   & 0 & 0 &0  & \frac{\zeta}{2}-\frac{2\beta}{3} & -\frac{\sqrt{2}}{3}\beta \\
   0     & 0 & 0 & 0 & -\frac{\sqrt{2}}{3}\beta  & -\zeta-\frac{\beta}{3} \\
\end{array} \right)
\end{displaymath}
These matrix elements are ordered following the state vectors $|SLJM\rangle$, or excluding $SL$ :
\begin{displaymath}
\hspace{2.5cm}{|J,M\rangle} =
\left( \begin{array}{c}
\frac{3}{2}, +\frac{3}{2} \\
\frac{3}{2}, -\frac{3}{2} \\
  \frac{3}{2}, +\frac{1}{2}    \\
  \frac{1}{2}, +\frac{1}{2}\\
 \frac{3}{2}, -\frac{1}{2}\\
\frac{1}{2}, -\frac{1}{2} \\
\end{array} \right)
\end{displaymath}
where the numbers $SL$ are in common to all states, e.g., $S=1/2$ and $L=1$ for the \ion{Ca}{ii}\,4p level.
The quadratic matrix elements are obtained following \citep{kem1975}:
\begin{equation}
    {\rm H_Q} = \frac{2}{3}\, \langle SLJM|C_0^{(0)}-C_0^{(2)}|SLJ^\prime M\rangle \ \langle r^2\rangle
\end{equation}
where $C_q^{(k)}$ are operators acting on the state vectors which we solved\footnote{We recovered the general expression for the matrix elements of the \ion{Ca}{ii}\,4p level as written in equation(8) of \citet{kem1975}, but we retained the more general formulation for the sign of the expression, i.e, $(-1)^{L+S+J+J^\prime-M}$, which would be applicable to other cases involving different $L$ values. However, as shown in the text, we could not recover all matrix elements as written in equations (9,11,12a,12b) of \citet{kem1975}.} following \citet{rac1942} and \citet{cow1981}. As demonstrated by \citet{kem1975} the non-zero matrix elements of the H$_{\rm Q}$ matrix follow the same tri-diagonal structure described above. Therefore, the following matrix elements are added to the Zeeman and linear Paschen-Back elements:

\begin{displaymath}
\mathbf{\rm H_Q= \langle r^2\rangle_{\rm 4p}\,\gamma B^2}
\left( \begin{array}{cccccc}
\frac{4}{5} & 0 & 0 & 0 & 0 &0  \\
   0     & \frac{4}{5} & 0 & 0 & 0 & 0 \\
   0     & 0 & \frac{8}{15} & \frac{2\sqrt{2}}{15} & 0 & 0 \\
  0      & 0 & \frac{2\sqrt{2}}{15} & \frac{2}{3} & 0 & 0 \\
     0   & 0 & 0 &0  & \frac{8}{15} & -\frac{2\sqrt{2}}{15} \\
   0     & 0 & 0 & 0 & -\frac{2\sqrt{2}}{15} & \frac{2}{3} \\
\end{array} \right)
\end{displaymath}
\begin{figure}
	\includegraphics[viewport= 10 170 580 690, clip, width=\columnwidth]{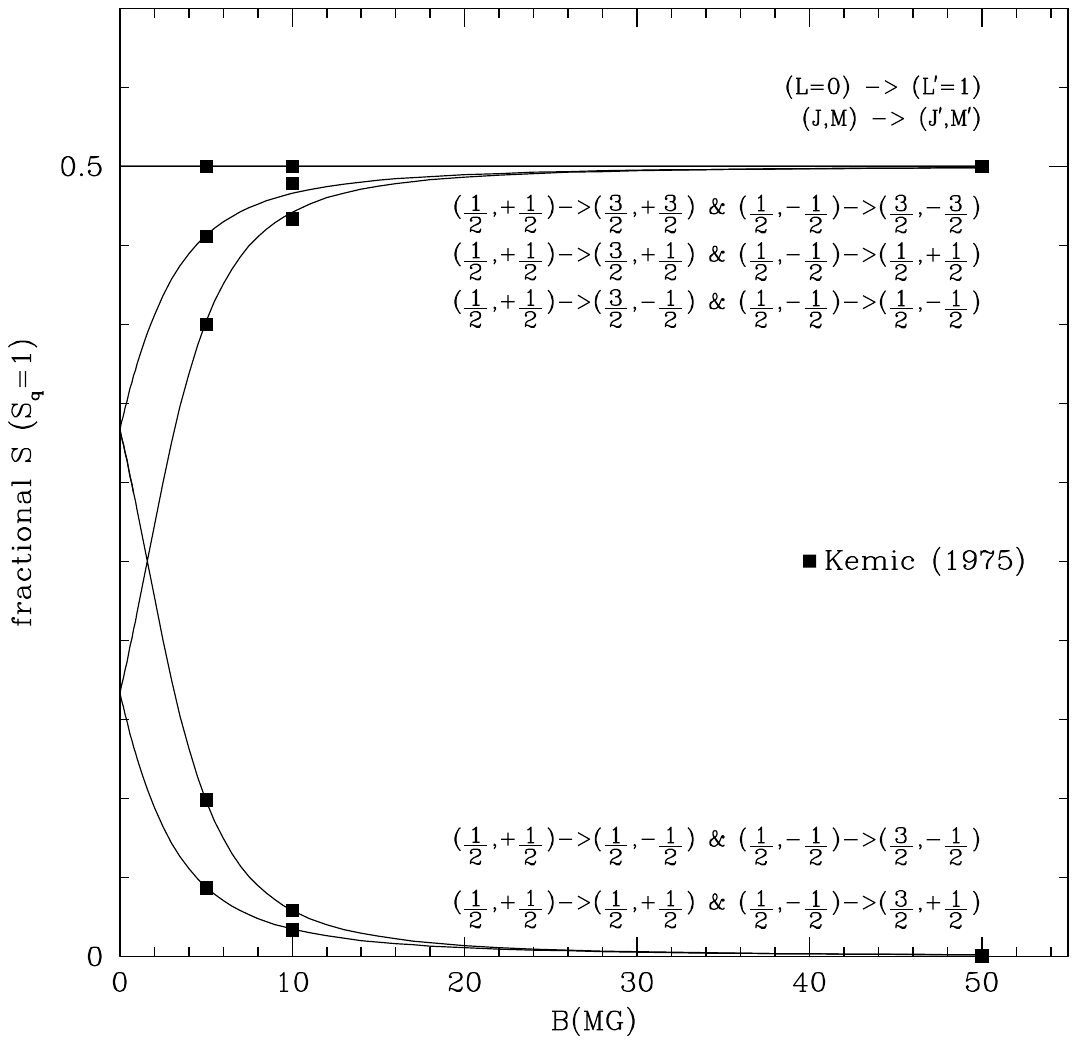}
    \caption{Fractional line strength as a function of magnetic field in the present work and compared to values tabulated in \citet{kem1975}. The curves are labelled, from top to bottom in order of appearance, with the corresponding line transition components. The top six line components survive into the full Paschen-Back regime while the lower four components vanish.}
    \label{fig_str}
\end{figure}

The angular part as described by \citet{kem1975} includes factors involving the 3-j and 6-j symbols described by \citet{rac1942} and more recently by \citet{cow1981} and \citet{lan2004}, and the matrix elements of spherical harmonics $\langle l||C^{(k)}||l^\prime\rangle$ \citep{rac1942,cow1981}. We solved the 3-j and 6-j symbols using fortran subroutines {\tt W3JS} and {\tt W6JS} supplied in \citet{lan2004}. The radial part involves the $\langle r^2\rangle$ expectation value (mean square radius) for a given configuration:
\begin{equation}
    \langle r^2\rangle_{nl} = \int r^2 P_{nl}^2(r) dr
\end{equation}
where $P_{\rm nl}$ is a normalized radial wave function.
Although distinct radial wave functions and $\langle r^2\rangle$ expectation values are suggested by \citet{kem1975} for the 4p levels P$_{3/2}$ and P$_{1/2}$ we adopted a single value for the entire upper level configuration, $\langle r^2\rangle_{\rm 4p}=22.04$~atomic units (a.u.), while we adopted $\langle r^2\rangle_{\rm 4s}=14.84$~a.u. for the lower level configuration. These and other required $\langle r^2\rangle$ values for various elements and configurations were computed by us using Cowan et al.'s fortran code {\tt RCN} following the Hartree-Fock (non-relativistic) scheme.  

We note that our H$_{\rm Q}$ coefficients differ by up to a factor of 2 from those of \citet{kem1975} which also appeared inconsistent with each other. The differences do not add up to large deviations in the resulting line wavelengths or strengths due to the relative dominance of the linear Paschen-Back effect in the range of magnetic field considered here ($\lesssim50$~MG).

\begin{figure}
	\includegraphics[viewport= 10 170 580 690, clip, width=\columnwidth]{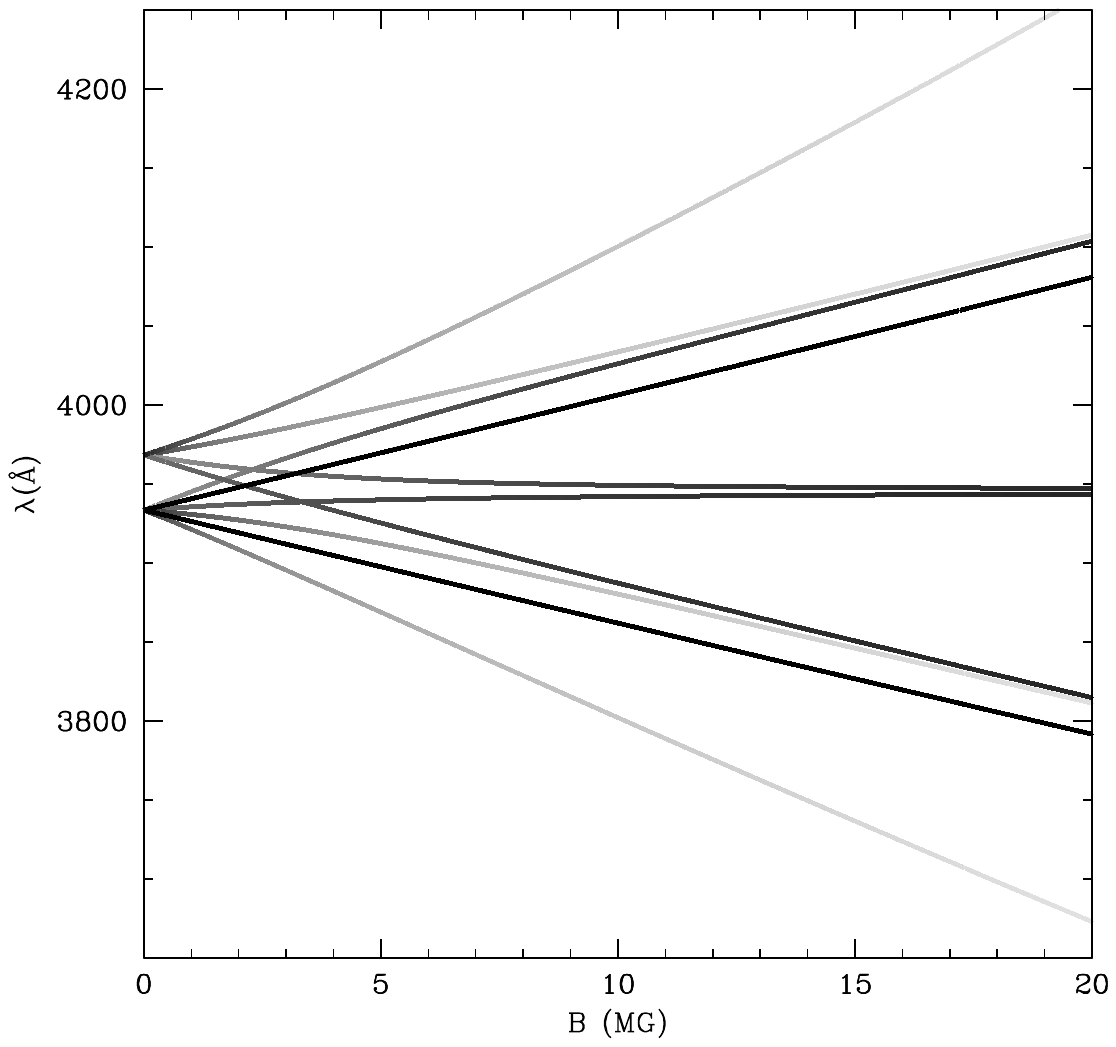}
    \caption{Shifted wavelength positions as a function of magnetic field of the Ca{\small II} H\&K doublet from zero field through the incomplete Paschen-Back regime and towards full Paschen-Back. The line strength is shown from full strength in black through declining strengths using lighter shades of grey towards the full Paschen-Back regime. At this stage, the polarization components $\pi$ and $\sigma_\pm$ retain doublet structures.}
    \label{fig_pba_cahk}
\end{figure}
Applying the selection rule $\Delta M=0,\pm1$ we recover 10 line components listed in the header of Table~\ref{tbl_CaHK_S}.
After solving for the eigenvalues and eigenvectors of the $\rm H_{so}+H_B+H_Q$ matrix using the subroutine {\tt jacobi} supplied in \citet{pre1986}, the wavelength of each component listed in Table~\ref{tbl_CaHK_W} was obtained using the calculated energy values:
\begin{equation}
\lambda_{\rm c} = \frac{10^8}{(E_{\rm 4p}+E_{\rm up,c})-(E_{\rm 4s}+E_{\rm low,c})}
\end{equation}
where $\lambda_{\rm c}$ is given in \AA\ and $c$ designates any of the individual line components $(JM)->(J^\prime M^\prime)$,
$E_{\rm up,c}$ is any eigenvalue of the upper level matrix, $E_{\rm 4s}=0$~cm$^{-1}$ and $E_{\rm 4p}=25340.10$~cm$^{-1}$ are the average configuration energy values recommended by NIST for the lower and upper level of the \ion{Ca}{ii} H\&K doublet. The energy $E_{\rm low,c}$ belongs to one of the two lower energy levels:
\begin{equation}
    E_{\rm low,c}=\pm\mu_0 B +\frac{2}{3} \langle r^2\rangle_{\rm 4s}\,\gamma B^2
\end{equation}
The wavelengths are converted from vacuum to air.

The method for calculating the strength of each line component is described by \citet{lan2004}.
The factors listed in Table~\ref{tbl_CaHK_S} are to be applied to the full line strength $S_{ik}$, e.g., $S_{ik}=27$ a.u. for the Ca{\rm II} H\&K doublet. Furthermore, the sum of these factors is normalized to unity within each polarization component ($q=0,\pm1$), i.e., $S_0=S_{-1}=S_{+1}=S_{ik}$, so that the sum of the polarization components weighted by the geometrical factors is equal to the total line strength $S_{ik}$:
\begin{equation}
\frac{1}{2}\sin^2{\theta}\,S_0
    +\frac{1}{4}(1+\cos^2{\theta})\, S_{-1}+\frac{1}{4}(1+\cos^2{\theta})\,S_{+1}=
    S_{ik}
\end{equation}
where $\theta$ is the angle between the local magnetic field and the line-of-sight. For the purpose of computing the line opacity, the line strength of each component
is converted into an oscillator strength following the relation:
\begin{equation}
    gf_{\rm c} = 303.754\,\frac{S_{\rm c}}{\lambda_{\rm c}}
\end{equation}
where $gf_{\rm c}$ is the product of the statistical weight and oscillator strength and $\lambda_{\rm c}$ is the shifted wavelength for a given component $(JM)->(J^\prime M^\prime)$ and provided in Table~\ref{tbl_CaHK_W}. Figure~\ref{fig_str} compares the results of our line strength calculations to values tabulated in \citet{kem1975}, but renormalized to $S_{q=0,\pm1}=1,$ for fields ranging from 0 to 50~MG showing an excellent agreement for all 10 line components although some tabulated values in \citet{kem1975} suffer from rounding errors.
Figure~\ref{fig_pba_cahk} shows our calculated line position and illustrated line strength for the Ca{\small II} H\&K doublet showing the transition from the anomalous Zeeman regime through the incomplete Paschen-Back regime.

\begin{table}
 \caption{Reference data for Paschen-Back calculations of the $^2$S and $^2$P$^{\rm o}$ terms. The configurations $i$ and $k$ designate the lower and upper levels, respectively, while $\zeta$ is the fine-structure energy separation constant, $S_{ik}$ is the total line strength, and $\langle r^2\rangle_{i}$  and $\langle r^2\rangle_{k}$ are the mean square radii for the lower and upper levels, respectively.}
 \label{atomic_data}
 \setlength\tabcolsep{0.235cm}
 \begin{tabular}{cccccc}
  \hline
  Line & configuration  & $\zeta$ ($^2$P$^{\rm o}$) & $S_{ik}$ & $\langle r^2\rangle_{i}$ &$\langle r^2\rangle_{k}$  \\
       &   $i$ - $k$          & (cm$^{-1}$)     &  (a.u.)  & (a.u.)  &  (a.u.) \\
  \hline
  \ion{Li}{i}~$\lambda$6707 & 2s\ -\ 2p  & 0.2267 & 33.0 & 17.47 & 27.07 \\
  \ion{Na}{i}~$\lambda$5891 & 3s\ -\ 3p  & 11.4642 & 37.3 & 19.66 & 40.25\\
  \ion{Al}{i}~$\lambda$3955$^a$ & 3p\ -\ 4s  & 74.707 & 9.05 & 13.70 & 64.41 \\
  \ion{K}{i}~$\lambda$7676  & 4s\ -\ 4p  & 38.474  & 50.46 & 28.49 & 53.02 \\
  \ion{Ca}{ii}~$\lambda$3945 & 4s\ -\ 4p & 148.593 & 27.0 &  14.84 & 22.04 \\
    \hline
 \end{tabular} \\
 \footnotesize{$^a$ Symmetric $^2$P$^{\rm o}$ - $^2$S terms.}
\end{table}

\subsection{Other line transitions in the incomplete Paschen-Back regime}\label{paschen2}

The formulation adopted for the \ion{Ca}{ii} H\&K doublet can be directly applied to other spectral lines with $^2$S-$^2$P$^{\rm o}$ terms such as the \ion{Na}{i}~$\lambda$5891 doublet. Table~\ref{atomic_data} lists the atomic data employed to extend the Paschen-Back calculations to these spectral lines.
We also employed the formalism of \citet{kem1975} to calculate the quadratic Paschen-Back effect in spectral lines with other types of configurations from $L=0$ to 4 (S, P, D, F, H) and total electronic spin from $S=0$ to 3 (Table~\ref{data_calcium}).
The linear Paschen-Back matrix elements applicable to these configurations were obtained following the method described by \citet{lan2004}.

In the presence of multiple terms in mixed configurations we computed, when available, the term-dependent $\langle r^2\rangle$ values following \citet{con1935}. For example, compare the $\langle r^2\rangle$ values in the calcium 3d4p $^3$F$^{\rm o}$, $^3$D$^{\rm o}$, and the $^3$P$^{\rm o}$ levels (Table~\ref{data_calcium}). In addition, $\langle r^2\rangle$ values proved strongly correlated with the calculated level energy values. Therefore we adjusted the correlation potential factor \citep{cow1981} to achieve a better match between the calculated energy values and the energy values tabulated at NIST \citep{kra2022}. 

In the main body of the text we reviewed the accuracy of the theory when confronting our models to high-resolution spectroscopy of the intermediate field white dwarf 2MASS~J0916$-$4215. The results were satisfactory in most cases, particularly the $^2$S - $^2{\rm P^o}$ lines. Comparing the line "center-of-gravity" with the full Paschen-Back calculations of \citet{hol2023} shows increasing deviations beyond 20~MG which for the spectral lines considered here would constitute a practical limit to the accuracy of the theory employed in our own calculations. However, the morphology of the \ion{Ca}{ii} lines in particular remains far from a simple triplet structure. Since they were computed in zero-field conditions, the mean square radii ($\langle r^2\rangle$) remain a major source of uncertainties in computations of the quadratic effect within the present scheme. In addition, failure of the LS-coupling in more complex or mixed electron configurations and uncertainties in Lande-g values also affect the accuracy of the linear Paschen-Back model predictions.

\begin{table*}
 \caption{Reference data for Paschen-Back calculations of the multiplet or singlet lower ($i$) and upper ($k$)  levels. Along with the electronic configurations and associated terms, we include the average lower ($E_i$) and upper ($E_k$) energy levels, the total line strength ($S_{ik}$), and the mean square radii for the lower and upper levels employed in the quadratic Paschen-Back calculations. We indicate whether a given singlet or multiplet is observed in the spectrum of \2mass0916.}
 \label{data_calcium}
 \begin{tabular}{ccccccccccccc}
  \hline
  Element & wavelength & \multicolumn{2}{c}{configuration} & \multicolumn{2}{c}{terms}  & $E_i$& $E_k$ & $S_{ik}$  & $\langle r^2\rangle_{i}$ &$\langle r^2\rangle_{k}$ & observed \\
    & (\AA)  &   $i$ & $k$ & $i$ & $k$ &  (cm$^{-1}$)  & (cm$^{-1}$)  &  (a.u.)  & (a.u.) &  (a.u.) & \\
  \hline
  \ion{Na}{i} & 8190 & 3p   & 3d    & $^2$P$^{\rm o}$ & $^2$D & 16967.63421 & 29172.8570 & 140. & 40.25 & 123.0 & Y \\
  \ion{Mg}{i} & 3835 & 3s3p & 3s3d & $^3$P$^{\rm o}$ & $^3$D & 21890.854 & 47957.042 & 67.3 & 19.23 & 89.78 & Y \\
            & 5178 & 3s3p & 3s4s  & $^3$P$^{\rm o}$ & $^3$S & 21890.854 & 41197.403 & 20.82 & 19.23 & 78.89 & Y \\
\ion{Si}{i} & 3905 & 3s$^2$3p$^2$ & 3s$^2$3p4s & $^1$S & $^1$P$^{\rm o}$ & 15394.370 & 40991.884 & 1.17 & 10.72 & 80.38 & N \\
\ion{Ca}{i} & 3637 & 4s4p   & 4s5d  & $^3$P$^{\rm o}$ & $^3$D & 15263.089 & 42745.620 & 13.0 & 29.29 & 565.6 & blend \\
            & 4226 & 4s$^2$ & 4s4p & $^1$S & $^1$P$^{\rm o}$ & 0.000 & 23652.304 & 24.4 & 19.31 & 43.01 & Y \\
  & 4300 & 4s4p & 4p$^2$ & $^3$P$^{\rm o}$ & $^3$P &15263.089 & 38507.751 & 64.0 & 29.29 & 29.33 & Y \\
            & 4445 & 4s4p & 4s4d & $^3$P$^{\rm o}$ & $^3$D & 15263.089 & 37753.738 & 57.0 & 29.29 & 147.2 & Y \\
            & 5266 & 3d4s & 3d4p & $^3$D & $^3$P$^{\rm o}$  & 20356.625 & 39337.750 & 39.0 & 22.41 & 44.52 & Y \\
            & 5592 & 3d4s & 3d4p & $^3$D & $^3$D$^{\rm o}$ & 20356.625 & 38232.442 & 71.0 & 22.41 & 43.79 & Y \\
            & 6142 & 4s4p & 4s5s  & $^3$P$^{\rm o}$ & $^3$S & 15263.089 & 31539.495 & 30.0 & 29.29 & 106.8 & Y  \\
            & 6460 & 3d4s & 3d4p & $^3$D & $^3$F$^{\rm o}$ & 20356.625 & 35831.203 & 150.0 & 22.41 & 40.39 & Y \\
\ion{Ca}{ii} & 8578 & 3d & 4p & $^2$D & $^2$P$^{\rm o}$ & 13686.60 & 25340.10 & 21.0 & 6.835 & 22.04 & Y \\ 
\ion{Ti}{i} & 3646 & 3d$^2$4s$^2$& 3d$^2$4s4p& a$^3$F & y$^3$G$^{\rm o}$ & 222.5141 & 27639.869 & 59.7 & 15.55 &  36.59  & blend \\
            & 3743 & 3d$^2$4s$^2$& 3d$^2$4s4p& a$^3$F& x $^3$F$^{\rm o}$&222.5141& 26928.504 & 33.0 & 15.55 & 35.64 & blend\\
            & 3991 & 3d$^2$4s$^2$& 3d$^2$4s4p& a$^3$F&y$^3$F$^{\rm o}$&222.5141&25267.742&35.2& 15.55 & 33.77 &  N \\
  & 4534 & 3d$^3$4s& 3d$^3$4p&a$^5$F& y$^5$F$^{\rm o}$& 6721.393 & 28767.276 & 172. & 17.93 & 36.07 & Y \\
            & 4997 & 3d$^3$4s& 3d$^3$4p&a$^5$F& y$^5$G$^{\rm o}$& 6721.393 & 26726.246 & 185. & 17.93 & 33.48 & Y \\
\ion{V}{i}   & 4392 & 3d$^4$4s&3d$^4$4p&a$^6$D& y$^6$F$^{\rm o}$&2296.5809&25056.952 &190.0& 16.37 & 32.10 & N \\
\ion{Cr}{i} & 3589 & 3d$^5$4s &3d$^4$4s4p&a$^7$S&y$^7$P$^{\rm o}$&0.000&27847.7433&73.0& 15.12 & 22.43 & Y \\
            & 4269 &3d$^5$4s &3d$^5$4p&a$^7$S&z$^7$P$^{\rm o}$&0.000&23415.1784& 25.3 & 15.12 & 29.17 & Y \\
            & 5206 &3d$^5$4s &3d$^5$4p&a$^5$S&z$^5$P$^{\rm o}$&7593.1484&26793.2864&53.2& 16.61 & 33.12 & Y \\
            & 4350 &3d$^4$4s$^2$ &3d$^4$4s4p &a$^5$D&z$^5$F$^{\rm o}$&8090.1903&31070.1022&17.0& 13.33 & 23.73 &  N \\
            & 5345 & 3d$^4$4s$^2$& 3d$^5$4p&a$^5$D&z$^5$P$^{\rm o}$&8090.1903&26793.2864&9.7& 13.33 & 33.12 & N \\
\ion{Mn}{i} & 4032 & 3d$^5$4s$^2$& 3d$^5$4s4p&a$^6$S & z$^6$P$^{\rm o}$& 0.000 & 24792.42& 9.9 & 12.48 & 23.01 &  N \\
\ion{Fe}{i} & 3456 & 3d$^6$4s$^2$ & 3d$^6$4s4p & a$^5$D& z$^5$P$^{\rm o}$ & 402.961 &29329.1731  & 7.62& 11.61 & 22.63 & Y \\
            & 3611 & 3d$^7$4s &3d$^7$4p & a$^5$F&z$^5$G$^{\rm o}$ & 7459.7517 &35143.4972 & 74.6 & 13.29 & 28.32 & Y \\
            & 3727 & 3d$^6$4s$^2$ &3d$^6$4s4p & a$^5$D& z$^5$F$^{\rm o}$& 402.961 & 27219.4636 & 13.7 & 11.61 & 21.69 & Y \\ 
            & 3750 & 3d$^7$4s &3d$^7$4p  & a$^5$F &y$^5$F$^{\rm o}$ & 7459.7517 & 34117.6793& 83.4 & 13.29 & 27.28 & Y \\
            & 3830 & 3d$^7$4s &3d$^7$4p & a$^3$F &y$^3$D$^{\rm o}$ & 12407.4028 & 38506.9925 & 55.4  & 14.06 & 32.76 & Y \\
            & 3838 & 3d$^7$4s &3d$^7$4p  & a$^5$F & y$^5$D$^{\rm o}$& 7459.7517 & 33503.7641& 45.7 & 13.29 & 26.72 & Y \\
            & 3882 & 3d$^6$4s$^2$ &3d$^6$4s4p & a$^5$D&z$^5$D$^{\rm o}$ & 402.961 &26150.7302 & 7.37 & 11.61 & 22.54 & Y \\
            & 4057 & 3d$^7$4s &3d$^7$4p  & a$^3$F &y$^3$F$^{\rm o}$ & 12407.4028 &37043.8386 &68.8  & 14.06 & 30.61 & Y \\
            & 4293 & 3d$^7$4s &3d$^7$4p  & a$^3$F &z$^3$G$^{\rm o}$ & 12407.4028 & 35690.1840 & 41.6 & 14.06 & 28.94 & N \\
            & 5059 & 3d$^7$4s &3d$^6$4s4p & a$^5$F & z$^5$F$^{\rm o}$& 7459.7517 & 27219.4636& 0.201 & 13.29 & 21.69 & N  \\
            & 5217 & 3d$^7$4s &3d$^6$4s4p &a$^3$F  & z$^3$D$^{\rm o}$& 12407.4028 & 31566.8014 & 3.55 & 14.06 & 22.96 & N  \\
            & 5348 & 3d$^7$4s & 3d$^6$4s4p& a$^5$F & z$^5$D$^{\rm o}$& 7459.7517 &26150.7302 &3.06 & 13.29 & 22.54 & N  \\
\ion{Ni}{i} & 3528 & 3d$^9$4s & 3d$^9$4p & $^3$D& $^3$P$^{\rm o}$& 731.457 & 29060.032& 21. & 11.94 & 25.14 & N \\
 & 3566 & 3d$^9$4s & 3d$^9$4p & $^1$D& $^1$D$^{\rm o}$& 3409.937 & 31441.635& 6.3 & 12.30 & 27.59 & N \\
            & 3619 & 3d$^9$4s & 3d$^9$4p & $^1$D& $^1$F$^{\rm o}$& 3409.937 & 31031.020& 11. & 12.30 & 27.09 & N \\
\ion{Sr}{i} & 4607 & 5s$^2$& 5s5p &$^1$S & $^1$P$^{\rm o}$ &0.000 &21698.452 & 29.1 & 22.54 & 50.18 & Y \\
 \hline
 \end{tabular} 
\end{table*}

\begin{table*}
 \caption{Fractional line strength $S$ for each Ca H\&K component as a function of the magnetic field strength in the incomplete Paschen-Back regime. Each component is labeled with the quantum numbers $(J,M)$ for the lower level ($^1$S) and $(J',M')$ for the upper level ($^2$P), and its membership to the H or K spectral line and grouped within the $\pi$ ($q=0$) or $\sigma_\pm$ ($q=\pm1$) components.}
 \label{tbl_CaHK_S}
 \begin{tabular}{cccccccccccccc}
  \hline
$q$   & \multicolumn{3}{c}{$-1$} & &\multicolumn{5}{c}{$0$} & & \multicolumn{3}{c}{$+1$} \\
   \cline{2-4} \cline{6-10} \cline {12-14}
 $M,M'$  & \multicolumn{1}{c}{$\frac{1}{2},\frac{3}{2}$} & \multicolumn{2}{c}{$-\frac{1}{2},\frac{1}{2}$} & & \multicolumn{2}{c}{$\frac{1}{2},\frac{1}{2}$} & & \multicolumn{2}{c}{$-\frac{1}{2},-\frac{1}{2}$} & & \multicolumn{2}{c}{$\frac{1}{2},-\frac{1}{2}$} & $-\frac{1}{2},-\frac{3}{2}$\\
   \cline{3-4} \cline{6-7} \cline{9-10} \cline{12-13}
$J,J'$   &$\frac{1}{2},\frac{3}{2}$ & $\frac{1}{2},\frac{3}{2}$ & $\frac{1}{2},\frac{1}{2}$& &$\frac{1}{2},\frac{3}{2}$ & $\frac{1}{2},\frac{1}{2}$& &$\frac{1}{2},\frac{3}{2}$ & $\frac{1}{2},\frac{1}{2}$& &$\frac{1}{2},\frac{3}{2}$ &$\frac{1}{2},\frac{1}{2}$ &  $\frac{1}{2},\frac{3}{2}$\\
Ca  & K & K & H & & K & H & & K & H & & K & H & K \\
               &    &    &    & &    &    & &    &    & &    &    &    \\
       $B$ (MG)           & \multicolumn{13}{c}{$S$} \\
\cline{2-14}   
  0.0 & 0.50000 & 0.16667 & 0.33333&& 0.33333 & 0.16667&& 0.33333 & 0.16667&& 0.16667 & 0.33333 & 0.50000 \\
  0.1 & 0.50000 & 0.16206 & 0.33794&& 0.33794 & 0.16206&& 0.32863 & 0.17137&& 0.17137 & 0.32863 & 0.50000 \\
  0.2 & 0.50000 & 0.15756 & 0.34244&& 0.34244 & 0.15756&& 0.32383 & 0.17617&& 0.17617 & 0.32383 & 0.50000 \\
  0.3 & 0.50000 & 0.15316 & 0.34684&& 0.34684 & 0.15316&& 0.31894 & 0.18106&& 0.18106 & 0.31894 & 0.50000 \\
  0.4 & 0.50000 & 0.14886 & 0.35114&& 0.35114 & 0.14886&& 0.31396 & 0.18604&& 0.18604 & 0.31396 & 0.50000 \\
  0.5 & 0.50000 & 0.14467 & 0.35533&& 0.35533 & 0.14467&& 0.30890 & 0.19110&& 0.19110 & 0.30890 & 0.50000 \\
  0.6 & 0.50000 & 0.14058 & 0.35942&& 0.35942 & 0.14058&& 0.30376 & 0.19624&& 0.19624 & 0.30376 & 0.50000 \\
  0.7 & 0.50000 & 0.13660 & 0.36340&& 0.36340 & 0.13660&& 0.29854 & 0.20146&& 0.20146 & 0.29854 & 0.50000 \\
  0.8 & 0.50000 & 0.13273 & 0.36727&& 0.36727 & 0.13273&& 0.29326 & 0.20674&& 0.20674 & 0.29326 & 0.50000 \\
  0.9 & 0.50000 & 0.12896 & 0.37104&& 0.37104 & 0.12896&& 0.28792 & 0.21208&& 0.21208 & 0.28792 & 0.50000 \\
  1.0 & 0.50000 & 0.12530 & 0.37470&& 0.37470 & 0.12530&& 0.28252 & 0.21748&& 0.21748 & 0.28252 & 0.50000 \\
  1.2 & 0.50000 & 0.11829 & 0.38171&& 0.38171 & 0.11829&& 0.27158 & 0.22842&& 0.22842 & 0.27158 & 0.50000 \\
  1.4 & 0.50000 & 0.11168 & 0.38832&& 0.38832 & 0.11168&& 0.26052 & 0.23948&& 0.23948 & 0.26052 & 0.50000 \\
  1.6 & 0.50000 & 0.10548 & 0.39452&& 0.39452 & 0.10548&& 0.24939 & 0.25061&& 0.25061 & 0.24939 & 0.50000 \\
  1.8 & 0.50000 & 0.09965 & 0.40035&& 0.40035 & 0.09965&& 0.23826 & 0.26174&& 0.26174 & 0.23826 & 0.50000 \\
  2.0 & 0.50000 & 0.09418 & 0.40582&& 0.40582 & 0.09418&& 0.22719 & 0.27281&& 0.27281 & 0.22719 & 0.50000 \\
  2.5 & 0.50000 & 0.08198 & 0.41802&& 0.41802 & 0.08198&& 0.20023 & 0.29977&& 0.29977 & 0.20023 & 0.50000 \\
  3.0 & 0.50000 & 0.07164 & 0.42836&& 0.42836 & 0.07164&& 0.17493 & 0.32507&& 0.32507 & 0.17493 & 0.50000 \\
  3.5 & 0.50000 & 0.06289 & 0.43711&& 0.43711 & 0.06289&& 0.15191 & 0.34809&& 0.34809 & 0.15191 & 0.50000 \\
  4.0 & 0.50000 & 0.05547 & 0.44453&& 0.44453 & 0.05547&& 0.13148 & 0.36852&& 0.36852 & 0.13148 & 0.50000 \\
  4.5 & 0.50000 & 0.04916 & 0.45084&& 0.45084 & 0.04916&& 0.11369 & 0.38631&& 0.38631 & 0.11369 & 0.50000 \\
  5.0 & 0.50000 & 0.04378 & 0.45622&& 0.45622 & 0.04378&& 0.09842 & 0.40158&& 0.40158 & 0.09842 & 0.50000 \\
  5.5 & 0.50000 & 0.03917 & 0.46083&& 0.46083 & 0.03917&& 0.08541 & 0.41459&& 0.41459 & 0.08541 & 0.50000 \\
  6.0 & 0.50000 & 0.03521 & 0.46479&& 0.46479 & 0.03521&& 0.07439 & 0.42561&& 0.42561 & 0.07439 & 0.50000 \\
  6.5 & 0.50000 & 0.03178 & 0.46822&& 0.46822 & 0.03178&& 0.06507 & 0.43493&& 0.43493 & 0.06507 & 0.50000 \\
  7.0 & 0.50000 & 0.02881 & 0.47119&& 0.47119 & 0.02881&& 0.05718 & 0.44282&& 0.44282 & 0.05718 & 0.50000 \\
  7.5 & 0.50000 & 0.02621 & 0.47379&& 0.47379 & 0.02621&& 0.05049 & 0.44951&& 0.44951 & 0.05049 & 0.50000 \\
  8.0 & 0.50000 & 0.02394 & 0.47606&& 0.47606 & 0.02394&& 0.04479 & 0.45521&& 0.45521 & 0.04479 & 0.50000 \\
  8.5 & 0.50000 & 0.02194 & 0.47806&& 0.47806 & 0.02194&& 0.03993 & 0.46007&& 0.46007 & 0.03993 & 0.50000 \\
  9.0 & 0.50000 & 0.02017 & 0.47983&& 0.47983 & 0.02017&& 0.03576 & 0.46424&& 0.46424 & 0.03576 & 0.50000 \\
  9.5 & 0.50000 & 0.01860 & 0.48140&& 0.48140 & 0.01860&& 0.03216 & 0.46784&& 0.46784 & 0.03216 & 0.50000 \\
 10.0 & 0.50000 & 0.01720 & 0.48280&& 0.48280 & 0.01720&& 0.02905 & 0.47095&& 0.47095 & 0.02905 & 0.50000 \\
 11.0 & 0.50000 & 0.01483 & 0.48517&& 0.48517 & 0.01483&& 0.02398 & 0.47602&& 0.47602 & 0.02398 & 0.50000 \\
 12.0 & 0.50000 & 0.01291 & 0.48709&& 0.48709 & 0.01291&& 0.02008 & 0.47992&& 0.47992 & 0.02008 & 0.50000 \\
 13.0 & 0.50000 & 0.01134 & 0.48866&& 0.48866 & 0.01134&& 0.01702 & 0.48298&& 0.48298 & 0.01702 & 0.50000 \\
 14.0 & 0.50000 & 0.01003 & 0.48997&& 0.48997 & 0.01003&& 0.01459 & 0.48541&& 0.48541 & 0.01459 & 0.50000 \\
 15.0 & 0.50000 & 0.00894 & 0.49106&& 0.49106 & 0.00894&& 0.01263 & 0.48737&& 0.48737 & 0.01263 & 0.50000 \\
 16.0 & 0.50000 & 0.00801 & 0.49199&& 0.49199 & 0.00801&& 0.01103 & 0.48897&& 0.48897 & 0.01103 & 0.50000 \\
 17.0 & 0.50000 & 0.00722 & 0.49278&& 0.49278 & 0.00722&& 0.00971 & 0.49029&& 0.49029 & 0.00971 & 0.50000 \\
 18.0 & 0.50000 & 0.00654 & 0.49346&& 0.49346 & 0.00654&& 0.00860 & 0.49140&& 0.49140 & 0.00860 & 0.50000 \\
 20.0 & 0.50000 & 0.00544 & 0.49456&& 0.49456 & 0.00544&& 0.00688 & 0.49312&& 0.49312 & 0.00688 & 0.50000 \\
 22.0 & 0.50000 & 0.00460 & 0.49540&& 0.49540 & 0.00460&& 0.00562 & 0.49438&& 0.49438 & 0.00562 & 0.50000 \\
 24.0 & 0.50000 & 0.00394 & 0.49606&& 0.49606 & 0.00394&& 0.00467 & 0.49533&& 0.49533 & 0.00467 & 0.50000 \\
 26.0 & 0.50000 & 0.00341 & 0.49659&& 0.49659 & 0.00341&& 0.00394 & 0.49606&& 0.49606 & 0.00394 & 0.50000 \\
 28.0 & 0.50000 & 0.00299 & 0.49701&& 0.49701 & 0.00299&& 0.00337 & 0.49663&& 0.49663 & 0.00337 & 0.50000 \\
 30.0 & 0.50000 & 0.00264 & 0.49736&& 0.49736 & 0.00264&& 0.00291 & 0.49709&& 0.49709 & 0.00291 & 0.50000 \\
 32.0 & 0.50000 & 0.00235 & 0.49765&& 0.49765 & 0.00235&& 0.00253 & 0.49747&& 0.49747 & 0.00253 & 0.50000 \\
 34.0 & 0.50000 & 0.00210 & 0.49790&& 0.49790 & 0.00210&& 0.00223 & 0.49777&& 0.49777 & 0.00223 & 0.50000 \\
 36.0 & 0.50000 & 0.00189 & 0.49811&& 0.49811 & 0.00189&& 0.00197 & 0.49803&& 0.49803 & 0.00197 & 0.50000 \\
 38.0 & 0.50000 & 0.00171 & 0.49829&& 0.49829 & 0.00171&& 0.00176 & 0.49824&& 0.49824 & 0.00176 & 0.50000 \\
 40.0 & 0.50000 & 0.00156 & 0.49844&& 0.49844 & 0.00156&& 0.00158 & 0.49842&& 0.49842 & 0.00158 & 0.50000 \\
 42.0 & 0.50000 & 0.00143 & 0.49857&& 0.49857 & 0.00143&& 0.00142 & 0.49858&& 0.49858 & 0.00142 & 0.50000 \\
 44.0 & 0.50000 & 0.00131 & 0.49869&& 0.49869 & 0.00131&& 0.00128 & 0.49872&& 0.49872 & 0.00128 & 0.50000 \\
 46.0 & 0.50000 & 0.00121 & 0.49879&& 0.49879 & 0.00121&& 0.00117 & 0.49883&& 0.49883 & 0.00117 & 0.50000 \\
 48.0 & 0.50000 & 0.00112 & 0.49888&& 0.49888 & 0.00112&& 0.00107 & 0.49893&& 0.49893 & 0.00107 & 0.50000 \\
 50.0 & 0.50000 & 0.00104 & 0.49896&& 0.49896 & 0.00104&& 0.00098 & 0.49902&& 0.49902 & 0.00098 & 0.50000 \\
\hline
\end{tabular}\\
\end{table*}

\begin{table*}
 \caption{Same as Table~\ref{tbl_CaHK_S} but for the air wavelength (in \AA) of each Ca H\&K component as a function of the magnetic field strength in the incomplete Paschen-Back regime and towards the full regime.}
 \label{tbl_CaHK_W}
 \begin{tabular}{cccccccccccccc}
  \hline 
$q$   & \multicolumn{3}{c}{$-1$} & &\multicolumn{5}{c}{$0$} & & \multicolumn{3}{c}{$+1$} \\
   \cline{2-4} \cline{6-10} \cline {12-14}
 $M,M'$  & \multicolumn{1}{c}{$\frac{1}{2},\frac{3}{2}$} & \multicolumn{2}{c}{$-\frac{1}{2},\frac{1}{2}$} & & \multicolumn{2}{c}{$\frac{1}{2},\frac{1}{2}$} & & \multicolumn{2}{c}{$-\frac{1}{2},-\frac{1}{2}$} & & \multicolumn{2}{c}{$\frac{1}{2},-\frac{1}{2}$} & $-\frac{1}{2},-\frac{3}{2}$\\
   \cline{3-4} \cline{6-7} \cline{9-10} \cline{12-13}
$J,J'$   &$\frac{1}{2},\frac{3}{2}$ & $\frac{1}{2},\frac{3}{2}$ & $\frac{1}{2},\frac{1}{2}$& &$\frac{1}{2},\frac{3}{2}$ & $\frac{1}{2},\frac{1}{2}$& &$\frac{1}{2},\frac{3}{2}$ & $\frac{1}{2},\frac{1}{2}$& &$\frac{1}{2},\frac{3}{2}$ &$\frac{1}{2},\frac{1}{2}$ &  $\frac{1}{2},\frac{3}{2}$\\
Ca  & K & K & H & & K & H & & K & H & & K & H & K \\
               &    &    &    & &    &    & &    &    & &    &    &    \\
       $B$ (MG)           & \multicolumn{13}{c}{$\lambda_{\rm air}$ (\AA)} \\
\cline{2-14}   
      0.00 &  3933.66 &  3933.66 &  3968.47 && 3933.66  & 3968.47 && 3933.66  & 3968.47 && 3933.66 &  3968.47 &  3933.66 \\
      0.10 &  3932.94 &  3932.46 &  3967.49 && 3933.90  & 3968.96 && 3933.42  & 3967.98 && 3934.87 &  3969.45 &  3934.39 \\
      0.20 &  3932.22 &  3931.24 &  3966.52 && 3934.13  & 3969.46 && 3933.17  & 3967.50 && 3936.06 &  3970.45 &  3935.11 \\
      0.30 &  3931.50 &  3930.02 &  3965.56 && 3934.36  & 3969.97 && 3932.91  & 3967.03 && 3937.25 &  3971.44 &  3935.83 \\
      0.40 &  3930.77 &  3928.80 &  3964.60 && 3934.57  & 3970.48 && 3932.65  & 3966.56 && 3938.43 &  3972.45 &  3936.56 \\
      0.50 &  3930.05 &  3927.57 &  3963.65 && 3934.79  & 3971.00 && 3932.37  & 3966.11 && 3939.61 &  3973.47 &  3937.28 \\
      0.60 &  3929.33 &  3926.34 &  3962.71 && 3934.99  & 3971.53 && 3932.09  & 3965.66 && 3940.78 &  3974.49 &  3938.00 \\
      0.70 &  3928.61 &  3925.09 &  3961.77 && 3935.19  & 3972.06 && 3931.81  & 3965.21 && 3941.94 &  3975.52 &  3938.73 \\
      0.80 &  3927.89 &  3923.85 &  3960.84 && 3935.39  & 3972.60 && 3931.51  & 3964.78 && 3943.09 &  3976.56 &  3939.45 \\
      0.90 &  3927.17 &  3922.60 &  3959.92 && 3935.58  & 3973.14 && 3931.21  & 3964.35 && 3944.24 &  3977.61 &  3940.17 \\
      1.00 &  3926.45 &  3921.35 &  3959.00 && 3935.76  & 3973.69 && 3930.90  & 3963.93 && 3945.38 &  3978.66 &  3940.90 \\
      1.20 &  3925.00 &  3918.82 &  3957.18 && 3936.11  & 3974.81 && 3930.25  & 3963.12 && 3947.64 &  3980.80 &  3942.35 \\
      1.40 &  3923.56 &  3916.29 &  3955.38 && 3936.45  & 3975.94 && 3929.58  & 3962.34 && 3949.87 &  3982.97 &  3943.80 \\
      1.60 &  3922.12 &  3913.73 &  3953.60 && 3936.76  & 3977.09 && 3928.87  & 3961.59 && 3952.07 &  3985.18 &  3945.25 \\
      1.80 &  3920.68 &  3911.17 &  3951.83 && 3937.06  & 3978.27 && 3928.13  & 3960.87 && 3954.24 &  3987.42 &  3946.70 \\
      2.00 &  3919.24 &  3908.59 &  3950.09 && 3937.34  & 3979.45 && 3927.36  & 3960.19 && 3956.38 &  3989.70 &  3948.15 \\
      2.50 &  3915.64 &  3902.09 &  3945.80 && 3937.97  & 3982.49 && 3925.29  & 3958.62 && 3961.60 &  3995.55 &  3951.78 \\
      3.00 &  3912.05 &  3895.54 &  3941.60 && 3938.53  & 3985.62 && 3923.03  & 3957.24 && 3966.63 &  4001.61 &  3955.41 \\
      3.50 &  3908.46 &  3888.93 &  3937.47 && 3939.01  & 3988.81 && 3920.60  & 3956.04 && 3971.50 &  4007.88 &  3959.05 \\
      4.00 &  3904.88 &  3882.30 &  3933.41 && 3939.44  & 3992.07 && 3918.01  & 3955.01 && 3976.22 &  4014.32 &  3962.69 \\
      4.50 &  3901.29 &  3875.64 &  3929.40 && 3939.82  & 3995.39 && 3915.29  & 3954.11 && 3980.80 &  4020.94 &  3966.33 \\
      5.00 &  3897.71 &  3868.96 &  3925.44 && 3940.16  & 3998.74 && 3912.45  & 3953.33 && 3985.26 &  4027.69 &  3969.98 \\
      5.50 &  3894.14 &  3862.28 &  3921.52 && 3940.46  & 4002.14 && 3909.51  & 3952.66 && 3989.63 &  4034.58 &  3973.63 \\
      6.00 &  3890.57 &  3855.58 &  3917.63 && 3940.73  & 4005.58 && 3906.48  & 3952.08 && 3993.92 &  4041.59 &  3977.29 \\
      6.50 &  3887.00 &  3848.89 &  3913.78 && 3940.98  & 4009.04 && 3903.39  & 3951.57 && 3998.13 &  4048.70 &  3980.95 \\
      7.00 &  3883.44 &  3842.19 &  3909.95 && 3941.20  & 4012.53 && 3900.23  & 3951.13 && 4002.29 &  4055.90 &  3984.61 \\
      7.50 &  3879.88 &  3835.50 &  3906.15 && 3941.40  & 4016.04 && 3897.03  & 3950.73 && 4006.40 &  4063.19 &  3988.28 \\
      8.00 &  3876.32 &  3828.82 &  3902.37 && 3941.59  & 4019.58 && 3893.78  & 3950.39 && 4010.46 &  4070.54 &  3991.95 \\
      8.50 &  3872.77 &  3822.14 &  3898.61 && 3941.76  & 4023.13 && 3890.50  & 3950.08 && 4014.49 &  4077.97 &  3995.62 \\
      9.00 &  3869.22 &  3815.48 &  3894.87 && 3941.91  & 4026.70 && 3887.19  & 3949.81 && 4018.50 &  4085.46 &  3999.30 \\
      9.50 &  3865.67 &  3808.83 &  3891.14 && 3942.06  & 4030.29 && 3883.85  & 3949.57 && 4022.47 &  4093.01 &  4002.98 \\
     10.00 &  3862.13 &  3802.19 &  3887.43 && 3942.20  & 4033.89 && 3880.49  & 3949.35 && 4026.43 &  4100.61 &  4006.66 \\
     11.00 &  3855.06 &  3788.97 &  3880.04 && 3942.44  & 4041.14 && 3873.73  & 3948.97 && 4034.29 &  4115.97 &  4014.04 \\
     12.00 &  3848.00 &  3775.80 &  3872.70 && 3942.66  & 4048.43 && 3866.92  & 3948.67 && 4042.11 &  4131.52 &  4021.44 \\
     13.00 &  3840.95 &  3762.71 &  3865.40 && 3942.85  & 4055.75 && 3860.07  & 3948.41 && 4049.88 &  4147.24 &  4028.85 \\
     14.00 &  3833.92 &  3749.69 &  3858.13 && 3943.02  & 4063.12 && 3853.19  & 3948.20 && 4057.64 &  4163.13 &  4036.28 \\
     15.00 &  3826.91 &  3736.74 &  3850.90 && 3943.18  & 4070.52 && 3846.30  & 3948.02 && 4065.37 &  4179.18 &  4043.72 \\
     16.00 &  3819.91 &  3723.88 &  3843.70 && 3943.32  & 4077.94 && 3839.39  & 3947.87 && 4073.09 &  4195.39 &  4051.17 \\
     17.00 &  3812.92 &  3711.09 &  3836.53 && 3943.46  & 4085.40 && 3832.48  & 3947.74 && 4080.81 &  4211.75 &  4058.64 \\
     18.00 &  3805.95 &  3698.38 &  3829.37 && 3943.58  & 4092.88 && 3825.56  & 3947.63 && 4088.53 &  4228.26 &  4066.13 \\
     20.00 &  3792.05 &  3673.20 &  3815.15 && 3943.82  & 4107.91 && 3811.74  & 3947.46 && 4103.96 &  4261.72 &  4081.14 \\
     22.00 &  3778.20 &  3648.35 &  3801.00 && 3944.02  & 4123.03 && 3797.92  & 3947.35 && 4119.41 &  4295.78 &  4096.22 \\
     24.00 &  3764.41 &  3623.82 &  3786.94 && 3944.22  & 4138.23 && 3784.13  & 3947.27 && 4134.88 &  4330.44 &  4111.35 \\
     26.00 &  3750.68 &  3599.60 &  3772.95 && 3944.41  & 4153.51 && 3770.38  & 3947.22 && 4150.40 &  4365.70 &  4126.54 \\
     28.00 &  3737.01 &  3575.71 &  3759.04 && 3944.58  & 4168.87 && 3756.66  & 3947.20 && 4165.95 &  4401.57 &  4141.80 \\
     30.00 &  3723.39 &  3552.12 &  3745.19 && 3944.76  & 4184.30 && 3742.99  & 3947.20 && 4181.55 &  4438.06 &  4157.11 \\
     32.00 &  3709.83 &  3528.85 &  3731.41 && 3944.93  & 4199.80 && 3729.36  & 3947.22 && 4197.20 &  4475.19 &  4172.49 \\
     34.00 &  3696.33 &  3505.87 &  3717.69 && 3945.10  & 4215.37 && 3715.78  & 3947.26 && 4212.91 &  4512.96 &  4187.92 \\
     36.00 &  3682.88 &  3483.19 &  3704.04 && 3945.27  & 4231.00 && 3702.25  & 3947.31 && 4228.66 &  4551.40 &  4203.42 \\
     38.00 &  3669.49 &  3460.81 &  3690.45 && 3945.44  & 4246.71 && 3688.77  & 3947.37 && 4244.47 &  4590.52 &  4218.97 \\
     40.00 &  3656.16 &  3438.71 &  3676.93 && 3945.61  & 4262.48 && 3675.34  & 3947.45 && 4260.34 &  4630.33 &  4234.59 \\
     42.00 &  3642.88 &  3416.89 &  3663.46 && 3945.79  & 4278.31 && 3661.96  & 3947.54 && 4276.26 &  4670.85 &  4250.26 \\
     44.00 &  3629.65 &  3395.35 &  3650.06 && 3945.97  & 4294.22 && 3648.63  & 3947.64 && 4292.24 &  4712.11 &  4266.00 \\
     46.00 &  3616.48 &  3374.09 &  3636.71 && 3946.16  & 4310.18 && 3635.36  & 3947.75 && 4308.28 &  4754.12 &  4281.80 \\
     48.00 &  3603.37 &  3353.09 &  3623.43 && 3946.34  & 4326.22 && 3622.14  & 3947.88 && 4324.38 &  4796.90 &  4297.66 \\
     50.00 &  3590.31 &  3332.35 &  3610.20 && 3946.53  & 4342.31 && 3608.97  & 3948.01 && 4340.53 &  4840.46 &  4313.57 \\
\hline
  \end{tabular}
\end{table*}

\newpage 
\section{Spectral atlas: MIKE and MagE spectra and spectral synthesis.}\label{atlas}

\begin{figure*}
	\includegraphics[viewport=115 180 490 690, clip, width=0.9\textwidth]{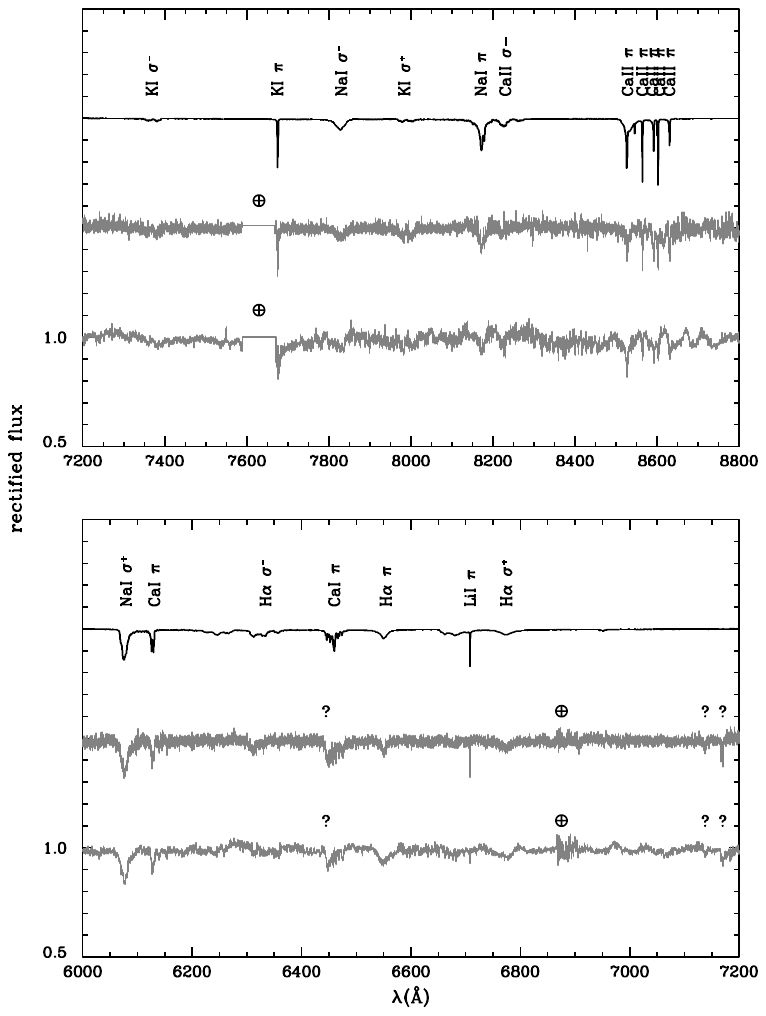}
    \caption{Model spectrum (top black line offset by +1.0), MIKE echelle spectrum (middle grey line offset by +0.5) and MagE echelle spectrum (bottom grey line) using air wavelengths. All spectral lines modelled in this work are marked with the corresponding element and polarization state ($\pi$ or $\sigma_\pm$). The question marks ("?") locate unknown features that would not belong to any of the elements identified in the spectrum. Regions affected by telluric absorption are marked with the $\earth$ symbol.}
    \label{fig_appen_a}
\end{figure*}

\begin{figure*}
	\includegraphics[viewport=115 180 490 690, clip, width=0.9\textwidth]{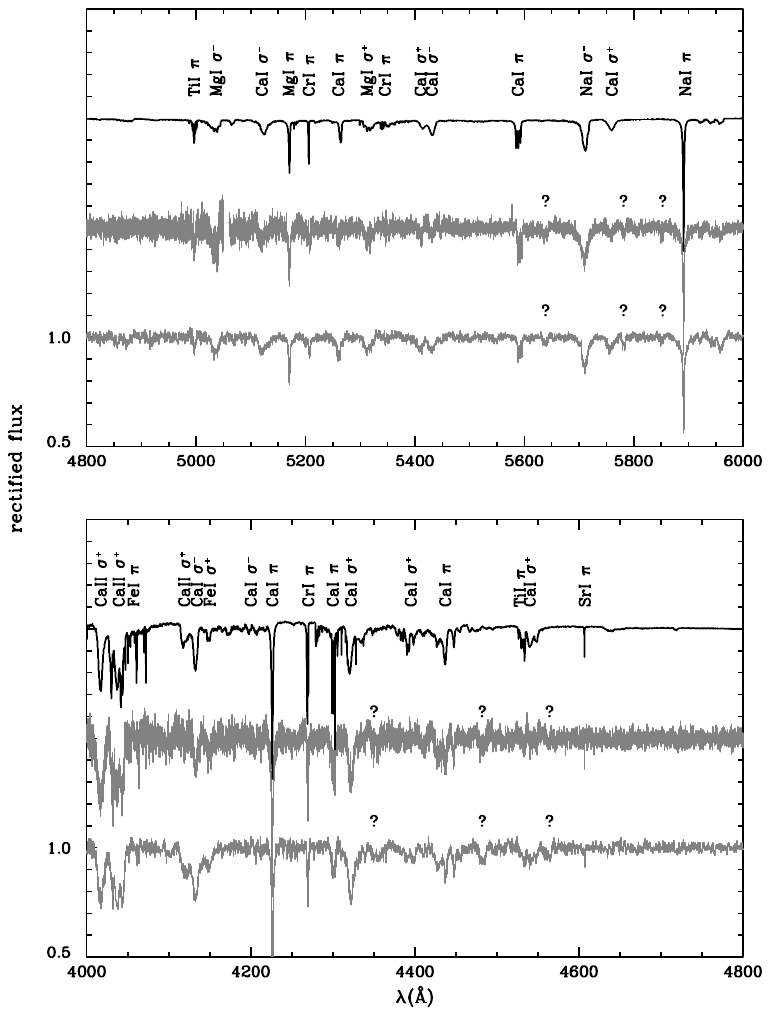}
    \caption{Same as Figure~\ref{fig_appen_a} and towards the blue end of the spectra.}
    \label{fig_appen_b}
\end{figure*}

The echelle spectra of \2mass0916\ and the spectral synthesis with a wavelength coverage from 4000 to 8800\AA\ are shown in four separate panels in  (Fig.~\ref{fig_appen_a} and Fig.~\ref{fig_appen_b}) complementing Figure~\ref{fig_spec}.

\bsp	% typesetting comment
\label{lastpage}
\end{document}